\documentclass[draft]{agujournal2019}
\usepackage{url} 
\usepackage{lineno}
\usepackage[inline]{trackchanges} 
\usepackage{soul}
\usepackage{tikz}
\newcommand*\circled[1]{\tikz[baseline=(char.base)]{
            \node[shape=circle,draw,inner sep=2pt] (char) {#1};}}

\draftfalse

\journalname{JGR: Space Physics}

\begin{document}

%
%

\title{Observations of an Electron-cold Ion Component Reconnection at the Edge of an Ion-scale Antiparallel Reconnection at the Dayside Magnetopause}

%
%




\authors{S. Q. Zhao\affil{1,2,3}, H. Zhang\affil{4}, Terry Z. Liu\affil{5}, Huirong Yan\affil{2,3}, C. J. Xiao\affil{1}, Mingzhe Liu\affil{6}, Q.-G. Zong\affil{7}, Xiaogang Wang\affil{8}, Mijie Shi\affil{9}, Shangchun Teng\affil{10}, Huizi Wang\affil{9}, R. Rankin\affil{11}, C. Pollock\affil{12}, G. Le\affil{13}}

\affiliation{1}{Fusion Simulation Center, State Key Laboratory of Nuclear Physics and Technology, School of Physics, Peking University, Beijing, China.}
\affiliation{2}{Deutsches Elektronen Synchrotron (DESY), Zeuthen, Germany.}
\affiliation{3}{Institut für Physik und Astronomie, Universität Potsdam, Potsdam, Germany.}
\affiliation{4}{Geophysical Institute, University of Alaska, Fairbanks, AK, USA.}
\affiliation{5}{Department of Earth, Planetary, and Space Sciences, University of California, Los Angeles, California, USA.}
\affiliation{6}{LESIA, Observatoire de Paris, Université PSL, CNRS, Sorbonne Université, Université de Paris, Meudon, France.}
\affiliation{7}{School of Earth and Space Sciences, Peking University, Beijing, China.}
\affiliation{8}{Department of Physics, Harbin Institute of Technology, Harbin, China.}
\affiliation{9}{Shandong Key Laboratory of Optical Astronomy and Solar-Terrestrial Environment, Institute of Space Sciences, Shandong University, Weihai, China.}
\affiliation{10}{CAS Key Laboratory of Geoscience Environment, School of Earth and Space Sciences, University of Science and Technology of China, Hefei, China.}
\affiliation{11}{Department of Physics, University of Alberta, Edmonton T6G 2G7 AB, Canada.}
\affiliation{12}{Denali Scientific, Fairbanks, AK, USA.}
\affiliation{13}{NASA Goddard Space Flight Center, Greenbelt, MD, USA}





\correspondingauthor{H. Zhang}{hzhang14@alaska.edu}




\begin{keypoints}
\item We report MMS observations of a multiscale, multi-type magnetopause reconnection in the presence of low-energy ions.
\item An electron-cold ion reconnection, dominated by low-energy ions and electrons, is located at the edge of an ion-scale reconnection.
\item The multiscale reconnection provides observational evidence for the simultaneous operation of antiparallel and component reconnection. 
\end{keypoints}

%
%

%
%


\begin{abstract}
Solar wind parameters play a dominant role in reconnection rate, which controls the solar wind-magnetosphere coupling efficiency at Earth’s magnetopause. Besides, low-energy ions from the ionosphere, frequently detected on the magnetospheric side of the magnetopause, also affect magnetic reconnection. However, the specific role of low-energy ions in reconnection is still an open question under active discussion. In the present work, we report in situ observations of a multiscale, multi-type magnetopause reconnection in the presence of low-energy ions using NASA’s Magnetospheric Multiscale data on 11 September 2015. This study divides ions into cold (10-500 eV) and hot (500-30000 eV) populations. The observations can be interpreted as a secondary reconnection dominated by electrons and cold ions (mainly in $XY_{GSE}$ plane) located at the edge of an ion-scale reconnection (mainly in $XZ_{GSE}$ plane). This analysis demonstrates a dominant role of cold ions in the secondary reconnection without hot ions’ response. Cold ions and electrons are accelerated and heated by the secondary process. The case study provides observational evidence for the simultaneous operation of antiparallel and component reconnection. Our results imply that the pre-accelerated and heated cold ions and electrons in the secondary reconnection may participate in the primary ion-scale reconnection affecting the solar wind-magnetopause coupling and the complicated magnetic field topology could affect the reconnection rate.
\end{abstract}

%
%

%


%
%
%
%

\section{Introduction}

Magnetic reconnection is a fundamental energy conversion process that efficiently releases magnetic energy into plasma energy via reconfiguration of magnetic topology in astrophysical and laboratory plasmas \cite{Dungey1961,Yamada2010}. Reconnection plays a crucial role in space weather phenomena, such as geomagnetic storms, energetic particle precipitation, and substorms \cite{Angelopoulos2008}. At the dayside magnetopause, reconnection is considered the main gateway wherein solar wind plasmas transfer mass, momentum, and energy to the magnetosphere. Solar wind parameters play a dominant role in reconnection rate, which controls the solar wind-magnetosphere coupling efficiency. Additionally, a mixture of ion compositions from the magnetosphere, magnetosheath, and ionosphere that appeared on the magnetospheric side of the magnetopause can also affect magnetic reconnection rate \cite{Borovsky2008,ToledoRedondo2021}. However, how low-energy ions affect reconnection is still not fully understood (e.g., \cite{Fuselier2016,Zhang2018}.

The low-energy ions convected to the magnetopause were detected ~15$\%$ of the time \cite{Darrouzet2008, Lee2016}, which can be distinguished by their energy, composition, and pitch angle distributions. There are two possible source regions: the plasmasphere and the ionosphere. 1) plasmaspheric ions are composed of $H^+$, $He^+$, and a small amount of $O^+$, reaching the magnetopause through plasmaspheric drainage plume or plasmaspheric wind \cite{Borovsky2008, Fuselier2016}. The plume is a sunward bulge driven by a convection electric field during high geomagnetic activities \cite{Moore2008}. The plasmaspheric wind is a continuous plasma flow originating from the plasmasphere during quiet and medium geomagnetic activities \cite{Walsh2013}. 2) ionospheric ions are composed of $H^+$, $O^+$, and a small amount of $He^+$, coming from ionospheric outflows or warm plasma cloak \cite{Chappell2008, Fuselier2016}. The ionospheric outflow with field-aligned pitch angle distributions is formed by the acceleration of the ambipolar electric field caused by thermal-electron pressure gradient or electron-ion separation. The warm plasma cloak is formed when ionospheric outflow is transported towards nightside and sent back into the dayside magnetosphere by convective drift, leading to $\sim90^\circ$ pitch angle distributions \cite{Chappell2008, Lee2016}.

When reaching the magnetopause, the dense cold ions may decrease the reconnection rate by increasing the plasma density and then decreasing the Alfvén speed at the reconnection site \cite{Walsh2013, Toledo-Redondo2015}. Low-energy ions introduce a new characteristic length scale determined by cold-ion gyroradius, creating a cold-ion diffusion region \cite{ToledoRedondo2021}. \citeA{Wang2015} suggested that cold ions might affect the reconnection rate only when entering the cold-ion diffusion region. When length scale L satisfies $R_{cold}< L< R_{hot}$, cold ions and electrons remain magnetized between hot-ion diffusion region edge and cold-ion diffusion region edge, where $R_{cold}$ and $R_{hot}$ are cold-ion and hot-ion gyroradius, respectively. In this case, cold ions can affect reconnection by modifying the generalized Ohm’s law \cite{Andre2016, Alm2018}. The cold-ion convection term decreases the measured electric field, and the enhanced density decreases the Hall electric field term. In contrast, some observations showed that low-energy ions play a minor role in reconnection. The low-energy ions maintained adiabatic motion and were entrained on the recently reconnected magnetic field by $\textbf{E}\times \textbf{B}$ drift \cite{Zhang2014}. In this process, low-energy ions were not involved in reconnection but were accelerated and heated by a pickup process \cite{Wang2014}. Therefore, the effects of low-energy ions on reconnection at the magnetopause are still debated. Additionally, the effects mentioned above require the entering of low-energy ions into the diffusion region, but whether the low-energy ions can still affect the reconnection outside the diffusion region is unknown. 

To further investigate the effects of low-energy ions on reconnection, we search for magnetopause reconnection in the presence of low-energy ions. The present work will show a representative case study of a multiscale, multi-type magnetopause reconnection involving low-energy ions on 11 September 2015 using the Magnetospheric Multiscale (MMS) measurements. This paper is organized as follows. In Section 2, we describe MMS measurements and the identifications of ion populations. Section 3 presents observational evidence for a smaller-scale current sheet located at the edge of an ion-scale reconnection. We define the smaller-scale current sheet as the electron-cold ion current sheet because its spatial scale is comparable to the electron and cold-ion inertial length. In Section 4, based on the measurements of four spacecraft, we discuss the possible interpretations of the electron-cold ion current sheet. In Section 5, we conclude our results.

\section{Methodology}

\subsection{Data and criteria}

The data used in this study are from the MMS mission, consisting of four identically instrumented spacecraft \cite{Burch2016a}. The magnetic field data with the sampling rate of 128 Hz are from the fluxgate magnetometer \cite{Russell2016}. The electric field data with the sampling rate of 8192 Hz are from the spin-plane and axial double-probes electric sensors \cite{Lindqvist2016,Ergun2016}. The plasma data are from the Fast Plasma Investigation instrument (FPI) \cite{Pollock2016} sampled at 30 ms for electrons and 150 ms for ions and the mass-resolved instrument Hot Plasma Composition Analyzer (HPCA) \cite{Young2016}. Moreover, we use the magnetic field and plasma data from the Wind mission \cite{Lepping1995, Lin1995} to verify the solar wind conditions. The AE-index data at 1 min resolution are from the NASA OMNIWeb.

We search for events that satisfy the following criteria: (1) spacecraft completes an entire magnetopause crossing where reconnection occurs. (2) low-energy ions near reconnection site, with a typical energy of a few to tens eV, identified from the ion energy flux spectrum. We find a new phenomenon: low-energy ions outside the diffusion region can introduce an electron-cold ion reconnecting current sheet at the edge of the ion-scale reconnection by involving a new length scale. We obtain four candidate events of such multiscale, multi-type reconnection involving low-energy ions. This study will present the most representative event (see supporting information for the other three events).

\subsection{Criteria for ion population identifications}

Figure 1 shows MMS1 observations of a magnetopause crossing from the magnetosphere to the magnetosheath during 07:55:27-07:55:40 UT on 11 September 2015. On the magnetospheric side of the magnetopause, three ion populations can be distinguished with their different energies, compositions, and pitch angle distributions: the magnetospheric ions, the ionospheric ions, and the magnetosheath ions. The identifications of ion populations used in this study are summarized below.

(1)	The magnetospheric ions are distinguished with high energy ($\sim$10 keV) in the ion energy flux spectrum (Figure 1k).

(2) The magnetosheath ions are distinguished with a typical energy of $\sim$1 keV and $\sim180^\circ$ pitch angles in FPI ion distributions after $\sim$07:55:30 UT (Figure 1p-1q). Besides, the population of moderate-energy $He^{++}$ with $0^\circ$ or $180^\circ$ pitch angles is another indication of the ions of magnetosheath origin in HPCA ion distributions (Figure 2h).

(3) In FPI ion distributions (Figure 1o-1q), a low-velocity and dense ion population mainly drifts at $\sim160$ $km s^{-1}$, roughly comparable to $\textbf{E}\times \textbf{B}$ drift speed $\sim210$ $km s^{-1}$ calculated from the electric field and magnetic field. HPCA ion distributions (Figure 2e-2h) show that the ion population moving in the direction perpendicular to the magnetic field is mainly composed of low-energy $H^+$ ($\sim$100 eV), moderate-energy $O^+$ ($\sim$1600 eV), and a small amount of $He^+$ during 07:55:22-07:55:32 UT. Due to the presence of large amounts of $O^+$, the effects of heavy ions on the ion energy cannot be ignored.

To further determine the composition and origin of the dense cold ions, we perform a Gaussian distribution fit with a one-dimensional cut of FPI ion perpendicular energy fluxes (pitch angle $\sim70^\circ-110^\circ$) \cite{Chen2004, Liang2015}. For FPI data, all ions are assumed to be protons \cite{Pollock2016}; however, ion species can be distinguished with their mass/charge ratio. Figure 3 shows the results of ion species separation analysis at 07:55:30.001 UT, based on MMS1 measurements. At this time interval, $H^+$ has an energy flux peak near $\frac{1}{2}m_{H^{+}}v_{i,perp}^2\sim 78.8$ $eV$ (purple vertical line), where $v_{i,perp}$ is the ion perpendicular velocity. $H^+$, $He^{++}$, $He^+$, and $O^+$ have a mass/charge ratio of 1, 2, 4, and 16, respectively. Therefore, if observed, $He^{++}$ should have an energy flux peak around $\frac{1}{4}m_{He^{++}}v_{i,perp}^2\sim m_{H^{+}}v_{i,perp}^2\sim 157.6$ $eV$ (red vertical line), $He^+$ should have an energy flux peak around $\frac{1}{2}m_{He^{+}}v_{i,perp}^2\sim 2m_{H^{+}}v_{i,perp}^2\sim 315.2$ $eV$ (blue vertical line), and $O^+$ should have a peak around $\frac{1}{2}m_{O^{+}}v_{i,perp}^2\sim 8m_{H^{+}}v_{i,perp}^2\sim1.26$ $keV$ (green vertical line). In Figure 3, we find a significant energy flux peak around the predicted $O^+$ energy, a weak energy flux enhancement around the predicted $He^+$ energy, but the trace of $He^{++}$ is not clear. Thus, we deduce that low-velocity and dense ion population is mainly composed of $H^+$, $O^+$, and, to a lesser extent, $He^+$, consistent with HPCA measurements. Since the observed ion composition is similar to that of ionospheric ions, the dense cold ions with $\sim90^\circ$ pitch angles likely originate from the warm plasma cloak.

To reduce the effects of heavy ions (especially for $O^+$) on the calculation, we divide the ions on the magnetospheric side of the magnetopause into cold (10-500 eV) and hot (500-30000 eV) ion populations with the critical energy of 500 eV (the horizontal dashed line in Figure 1k, black circles in Figure 2e-2h). The cold ion population is dominantly composed of $H^+$. The hot ion population is a mixture of magnetospheric ions, magnetosheath ions, and moderate-energy ionospheric heavy ions.

We compare bulk velocities and densities in cold and hot ion energy ranges from FPI and HPCA measurements to examine the validity of 500 eV as the critical energy. Since ions of different origins mix and cannot be distinguished after $\sim$07:55:32.4 UT, we only focus on the results at $\sim$07:55:22 UT. In Figure 2a-2d, solid curves denote FPI data; stars, pluses, and squares denote HPCA measurements of $H^+$, $O^+$, and $He^+$, respectively. The cold-ion bulk velocities observed by FPI are in good agreement with $H^+$ velocities (Figure 2b). The hot-ion bulk velocities observed by FPI are a bit closer to $O^+$/$He^+$ velocities (Figure 2c). Moreover, cold-ion density (black curve) is comparable to $H^+$ density (red stars), whereas hot-ion density (blue curve) is comparable to $O^+$ density (green pluses) in Figure 2d. Therefore, it is reasonable to employ 500 eV to distinguish $H^+$ from the heavy ions.

\section{Observations}

We investigate a multiscale, multi-type reconnection at the dayside magnetopause from 07:55:27-07:55:40 UT on 11 September 2015 (Figure 1). In mission phase 1a, the MMS spacecraft cruise across the dayside magnetopause from dusk to dawn \cite{Fuselier2016}. In this event, the spacecraft are located at $\sim$[5.5,6.0,0.3]  $R_E$ in Geocentric Solar Ecliptic (GSE) coordinates around 07:55:30 UT. Figure 4a-4d show the solar wind conditions, where the average solar wind speed $V_{X,GSE}$ is $\sim$540 $km s^{-1}$ from the Wind spacecraft at $[253.7,62.5,10.4]_{GSE}$. Therefore, the solar wind data should be shifted by $\sim$49 minutes to match MMS measurements. Figure 4a presents that $B_{Z,GSE}$ turns negative after 07:05:00 UT, indicating the southward interplanetary magnetic field. Figure 4d shows that the dynamic pressure fluctuates around 4.5 nPa. The AE-index data increase rapidly around 07:38:00 UT (Figure 4e), suggesting a strong magnetospheric convection condition.

\subsection{Overview of the ion-scale reconnection}

We first investigate the ion-scale reconnection marked by the red bar at the top of Figure 1a. Data are projected into local magnetic normal coordinates of the ion-scale current sheet ($[LMN]_{ion-scale}$) obtained from the minimum variance analysis of the magnetic field \cite{Sonnerup1998} during 07:55:28-07:55:40 UT. The eigenvectors are $\textbf{L}_{ion-scale}= [0.31, -0.22, 0.92]_{GSE}$, $\textbf{M}_{ion-scale}= [0.25, -0.92, -0.30]_{GSE}$, and $\textbf{N}_{ion-scale}= [0.92, 0.32, -0.23]_{GSE}$, respectively, where $\textbf{L}_{ion-scale}$ is along the reconnecting field, $\textbf{M}_{ion-scale}$ is in the guide field direction, and $\textbf{N}_{ion-scale}$ is along the magnetopause normal direction. Thus, the ion-scale reconnection roughly operates in $XZ_{GSE}$ plane.

MMS1 completes an outbound transition from the magnetosphere to the magnetosheath as characterized by the positive-to-negative magnetic field $B_{L,ion-scale}$ reversal across the magnetic magnitude minimum (Figure 1e), the increasing densities (Figure 1j), and the increasing differential energy fluxes of magnetosheath ions and electrons (Figure 1k and 1n). The magnetic shear angle between magnetospheric and magnetosheath magnetic field is $\sim153^\circ$, indicating an antiparallel reconnection. In Figure 1f, the high-speed southward ion flows ($V_{i,L,ion-scale}<0$) suggest that the spacecraft is located south of the magnetic X-line. It is an asymmetric reconnection because the magnetic field ratio is $\sim$5:4 (Figure 1e), and the density ratio is ~1:5 (Figure 1j). For the asymmetric reconnection, the maximum ion outflow $V_{i,L,ion-scale}\sim-$483 $km s^{-1}$ is in reasonable agreement with the hybrid Alfvén speed $V_{Ah}\sim-$399 $km s^{-1}$ derived from the parameters $B_{ph}\sim$101 nT, $B_{sh}\sim-$80 nT, $n_{ph}\sim8$ $cm^{-3}$, and $n_{sh}\sim37$ $cm^{-3}$ \cite{Cassak2007}. 

Figure 1e shows that a bipolar $B_{M,ion-scale}$ superposes on the background guide field of $\sim-$26.3 nT. The profile of the Hall quadrupolar structure is asymmetric, where $B_{M,ion-scale}$ with positive polarity is located at the separatrix layer of the ion-scale reconnection, whereas $B_{M,ion-scale}$ with negative polarity almost spans the whole outflow region, consistent with previous observations of asymmetric reconnection \cite{Tanaka2008, Wang2017}. MMS1 encounters the central plane of the ion-scale current sheet twice, seen from $B_{L,ion-scale}$ approaching zero twice ($\sim$07:55:34.0 UT and $\sim$07:55:36.5 UT, Figure 1e), most likely due to the back-and-forth motion of the magnetopause. The reconnection geometry and corresponding spacecraft trajectory are sketched in Figure 5a. 

Due to their different mass and energy, charged particles (electrons, cold ions, and hot ions) move along the magnetic field at different speeds and convect with the magnetic field at the same $\textbf{E}\times \textbf{B}$ speed. As a result, three types of flow boundaries between reconnection inflows and outflows on the magnetospheric side are presented, which are electron boundary (vertical dashed line $\circled{1}$ $\sim$07:55:28 UT), hot ion boundary (vertical dashed line $\circled{2}$ $\sim$07:55:29.9 UT), and cold ion boundary (vertical dashed line $\circled{3}$ $\sim$07:55:32.4 UT) in Figure 1. The electron boundary is characterized by the first detection of magnetosheath electrons with typical energy of hundreds of eV in the electron energy flux spectrum in Figure 1n \cite{Engwall2009}. The hot ion boundary is characterized by the first detection of magnetosheath ions with $\sim180^\circ$ pitch angles and typical energy of $\sim$1 keV, and the sharp increase in hot-ion bulk flow ($V_{i,hot,L,ion-scale}$) in Figure 1m and 1h \cite{Engwall2009}. We define the cold ion boundary as the dividing line between cold-ion inflow and outflow and characterize it by the sharp increase in cold-ion bulk flow ($V_{i,cold,L,ion-scale}$; Figure 1g).

Remarkably, an electron-cold ion scale current, which is mainly produced by high-speed electron flows, is observed at the edge of the ion-scale reconnection marked by the yellow-shaded region (Figure 1d). In contrast to ion outflows along $\textbf{L}_{ion-scale}$ direction, cold-ion jets, relative to the background velocities, are mainly along $\textbf{M}_{ion-scale}$ direction (Figure 1g), suggesting that cold ions experience a distinct dynamic process from hot ions. 

\subsection{Overview of the electron-cold ion current sheet}

To explore more explicit properties of the electron-cold ion current sheet, we zoom in to smaller timescales (Figure 6 and Figure 7). Due to large spacecraft relative separations ($\sim$210 km), data are projected into their respective local magnetic normal coordinates $[LMN]_{e-ci-*}$, where the subscript * represents spacecraft M1, M2, M3, and M4, respectively. $[LMN]_{e-ci-*}$ are derived from the hybrid minimum-variance analysis under low-magnetic-shear current sheet conditions \cite{Gosling2013}. For MMS1 observations, the current sheet normal is calculated by $\textbf{N}_{e-ci-M1}=\textbf{B}_{1}^{'}\times \textbf{B}_{2}^{'}/|\textbf{B}_{1}^{'}\times \textbf{B}_{2}^{'}|$, where $\textbf{B}_{1}^{'}$ is the average magnetic field over 07:55:27.547 ($t_{1}-2s$) to 07:55:29.547 ($t_{1}$) UT, and $\textbf{B}_{2}^{'}$ is the magnetic field minimum at the other edge of the current sheet at 07:55:30.374 ($t_{2}$) UT (since it is not clear whether MMS1 completes the entire crossing). The guide field direction is calculated by $\textbf{M}_{e-ci-M1}=\textbf{N}_{e-ci-M1}\times \textbf{L}_{e-ci-M1}^{'}$, where $\textbf{L}_{e-ci-M1}^{'}$ is the maximum variance direction of the magnetic field from $t_{1}$ to $t_{2}$. The reconnecting field direction $\textbf{L}_{e-ci-M1}=\textbf{M}_{e-ci-M1}\times \textbf{N}_{e-ci-M1}$ completes the coordinate system. The eigenvectors in $[LMN]_{e-ci-M2}$, $[LMN]_{e-ci-M3}$, and $[LMN]_{e-ci-M4}$ coordinates are obtained with the same method. The zoom-in plot (Figure 6) shows that MMS1 and MMS2 traverse the electron-cold ion current sheet in sequence at $\sim$07:55:30.4 UT. Subsequently, the zoom-in plot (Figure 7) shows that MMS3 and MMS4 traverse the same current sheet at $\sim$07:55:33.4 UT. 

Figure 6d shows that the field-aligned flow of electrons ($V_{e,M,e-ci-M1}$) reaches a maximum of $\sim$1872 $km s^{-1}$, which is responsible for the current along $\textbf{M}_{e-ci-M1}$ direction and roughly consistent with the gradient of $B_{L,e-ci-M1}$ across the current sheet. The electron jet in the reconnecting field direction ($V_{e,L,e-ci-M1}$) peaks at $-$558 $km s^{-1}$, significantly exceeding the local cold-ion Alfven speed $V_{A,i,cold}=B/\sqrt{4\pi m_{i}n_{i,cold}}\sim364$ $km s^{-1}$, where $n_{i,cold}$ refers to cold-ion density. Since $V_{e,L,e-ci-M1}$ and $V_{e,M,e-ci-M1}$ peak at different times, different components of electron jets may be formed by different physical processes rather than the projections of one electron jet in different directions. The other three spacecraft observed similar super-cold-ion-Alfvénic electron jets (Figure 6o, 7d, and 7o). 

To investigate whether the cold-ion bulk velocity is associated with a reconnecting ion outflow, we perform pressure anisotropy-weighted Walén tests \cite{Paschmann2000} in $\textbf{L}_{e-ci-*}$ direction. The Walén relation is expressed as $V=V_{0}+\triangle V$, where $\triangle V=\pm [B(1-\alpha)-B_{0}(1-\alpha)]/[\rho_{i,cold} \mu_{0}(1-\alpha) ]^{1/2}$, $B_{0}$ and $V_{0}$ are the magnetic field and velocity at two edges of the current sheet, $\alpha=(P_{\parallel}-P_{\perp}) \mu_{0}/B^{2}$ is the pressure anisotropy, $rho_{i,cold}$ is the cold-ion mass density, and $mu_{0}$ is the vacuum permeability. ‘+’ and ‘$-$‘ stand for in-phase and anti-phase relationship between the magnetic field and ion bulk velocity, respectively. MMS1, MMS2, and MMS4 observe that cold-ion bulk velocities $V_{i,cold,L,e-ci-M1}$, $V_{i,cold,L,e-ci-M2}$, and $V_{i,cold,L,e-ci-M4}$ (blue curves) are roughly consistent with respective Walen relation (black dashed curve) across the electron-cold ion current sheet, respectively (Figure 6e, 6p, and 7e), indicating the reconnecting ion outflows. Although the correlation between $V_{i,cold,L,e-ci-M3}$ and the Walén relation is not significant as observations from the other three spacecraft, the cold-ion bulk velocities observed by MMS3 still present a good correlation tendency in Figure 7p.

Similarly, to investigate whether the hot-ion bulk velocity is associated with a reconnecting ion outflow in $\textbf{L}_{e-ci-*}$ direction, we perform pressure anisotropy-weighted Walén tests on the hot ion population. Four spacecraft observe that the hot-ion bulk velocities $V_{i,hot,L,e-ci-*}$  are almost constant (Figure 6g, 6r, 7g, and 7r) and differ significantly from respective Walén relation (not shown) across the current sheet. Therefore, there is no clear evidence for hot-ion outflows in $\textbf{L}_{e-ci-*}$ direction.

To further confirm the relationship between the cold-ion bulk velocity and reconnecting outflows, we perform Walén tests on electron-cold ion current sheet with the cold ion population in L-, M-, and N- components of $[LMN]_{e-ci-*}$ coordinates \cite{Sonnerup1987}. Since four spacecraft show similar observations, we only present MMS1 results to simplify the expression. Figure 8a and 8b show the DeHoffmann-Teller (HT) analysis of intervals during 07:55:29.30-07:55:30.30 UT and 07:55:30.30-07:55:31.30 UT, respectively. The cold-ion convection electric fields ($-\textbf{V}_{i,cold}\times \textbf{B}$) are consistent with $-\textbf{V}_{HT}\times \textbf{B}$, which can be seen from the slopes $\sim$1 and correlation coefficients $\sim$1. Figure 8c and 8d show that cold-ion bulk velocities in the HT frame are linearly positively correlated with $V_{A,i,cold}$ (absolute values of slopes $\sim$1, and linear correlation coefficients are 0.97 and 0.99, respectively), indicating the appearance of a reconnecting cold-ion outflow on each side of the electron-cold ion current sheet.

We investigate the frozen-in condition by comparing $L_{e-ci-*}$ components of electron (blue), cold-ion (green), and hot-ion (red) perpendicular velocities to $[\textbf{E}\times \textbf{B}/B^{2} ]_{L_{e-ci-*}}$ (black) in Figure 6h, 6s, 7h, and 7s. The electron perpendicular velocities observed by MMS1 (Figure 6h) and MMS4 (Figure 7h) are roughly consistent with corresponding  $[\textbf{E}\times \textbf{B}/B^{2} ]_{L_{e-ci-*}}$  with the exception of short periods (around 07:55:30.3 UT and around 07:55:33.8 UT, respectively). In contrast, MMS2 and MMS3 observe that electron perpendicular velocities are always consistent with  $[\textbf{E}\times \textbf{B}/B^{2} ]_{L_{e-ci-*}}$ across the electron-cold ion current sheet. It suggests that electrons observed by MMS2 and MMS3 remain magnetized, but those observed by MMS1 and MMS4 experience a short-interval demagnetization. MMS1, MMS2, and MMS4 observe that cold-ion perpendicular velocities are comparable to  $[\textbf{E}\times \textbf{B}/B^{2} ]_{L_{e-ci-*}}$, indicating that the cold ion population remains coupled with the magnetic field. However, MMS3 observes that cold ions decouple from the magnetic field at ~07:55:33.7 UT (Figure 7p). All spacecraft observe that hot-ion perpendicular velocities differ strikingly from  $[\textbf{E}\times \textbf{B}/B^{2} ]_{L_{e-ci-*}}$, suggesting that hot ions are always demagnetized across the current sheet.

Figures 6j, 6u, 7j, and 7u show the energy conversion rates $\textbf{J}\cdot \textbf{E}^{'}$, which are dominated by parallel components, where $\textbf{J}=en_{e}(\textbf{v}_{i}-\textbf{v}_{e})$ is the current density, $\textbf{E}^{'}=\textbf{E}+\textbf{v}_{e}\times \textbf{B}$, $n_{e}$ is the electron density, and e is the elementary charge. MMS4 observes a positive $\textbf{J}\cdot \textbf{E}^{'}$ peak (Figure 7j), coinciding with the electron jet peak (Figure 7d), parallel electric field peak (Figure 7i), and parallel electron temperature peak (Figure 7k). The positive $\textbf{J}\cdot \textbf{E}^{'}$ represents a dissipative process that converts magnetic energy to plasma energy \cite{Burch2016b}, consistent with enhancing the parallel electron temperature in Figure 7k. Figure 6j shows that MMS1 observes a bipolar $\textbf{J}\cdot \textbf{E}^{'}$ fluctuation, in agreement with previous simulations and observations near the outer electron diffusion region \cite{Chen2016, Hwang2017}. In Figure 6u, MMS2 observes a $\textbf{J}\cdot \textbf{E}^{'}$, fluctuation with the longer negative part. In Figure 7u, MMS3 observes $\textbf{J}\cdot \textbf{E}^{'}$, with large-amplitude fluctuations.

\section{Discussion}
\subsection{One possible interpretation of the electron-cold ion current sheet: a multiscale, multi-type reconnection scenario}

Four spacecraft detect super-cold-ion-Alfvénic electron jets $V_{e,L,e-ci-*}$ (Figure 6d, 6o, 7d, and 7o) and cold-ion-Alfvénic cold-ion jets $V_{i,cold,L,e-ci-*}$ (Figure 6e, 6p, 7e, and 7o), accompanied by the apparent magnetic field $B_{L,e-ci-*}$ reversals (Figure 6b, 6m, 7b, and 7m), energy conversion (Figure 6j, 6u, 7j, and 7u), and electron temperature enhancements (Figure 6k, 6v, 7k, and 7v). Therefore, one possible interpretation of the electron-cold ion current sheet is that a secondary reconnection operates in the context of an ion-scale reconnection. 

The differences in observations from four spacecraft are attributed to their relative position from the X-line. We deduce that MMS4 is closest to the electron dissipation region due to the significant energy dissipative process ($\textbf{J}\cdot \textbf{E}^{'}>0$; Figure 7j). MMS1 is located a little further than MMS4, with a bipolar $\textbf{J}\cdot \textbf{E}^{'}$ fluctuation. Figure 6h shows that $V_{\perp,L,e-ci-M1}$ is faster than $[\textbf{E}\times \textbf{B}]_{L,e-ci-M1}$ around 07:55:30.3 UT. Thus, the negative part of $\textbf{J}\cdot \textbf{E}^{'}$ could be produced by a process in which electron outflows outrun and lash against the moving magnetic field, and then electrons do work on the magnetic field. Moreover, MMS1 observes a bipolar structure of the parallel electric field across the electron jet region, which is considered as a signal of the presence of a reconnection site nearby \cite{Lapenta2011}. The bipolar structure of the electric field may be generated by streaming instabilities resulting from the field-aligned electron velocities significantly exceeding the electron thermal velocity \cite{Goldman2008}. MMS2 with a longer-negative part of $\textbf{J}\cdot \textbf{E}^{'}$ may be even further away from the X-line than MMS1. The electrons are decelerated from the electron diffusion region edge to the cold-ion diffusion region. The longer-negative $\textbf{J}\cdot \textbf{E}^{'}$ leads to a flow deceleration in a more extended period, consistent with slower electron outflows ($|V_{e,L,e-ci-M2}|\sim384$ $kms^{-1}<|V_{e,L,e-ci-M1}|\sim558$ $kms^{-1}$). MMS3 observes relatively slow cold-ion outflow ($|V_{i,cold,L,e-ci-M3}-V_{0}|\leq 140$ $kms^{-1}$). Besides, MMS3 observations show the deviation between the cold-ion outflow and Walén relation (Figure 7p) and the parallel electric field with large-amplitude and high-frequency fluctuations (Figure 7t). This can be explained by the fact that the magnetic field is possibly affected by plasma waves. Therefore, we deduce that MMS3 is located furthest from the X-line. These conjectures are consistent with their actual relative positions between MMS1 (3) and MMS2 (4), as shown in Figure 5b (5c).

The electron-cold ion current sheet geometry is invariant in a short interval due to the small angles between $\textbf{N}_{e-ci-M1}$ and $\textbf{N}_{e-ci-M2}$ (around $5^\circ$ at ~07:55:30.4 UT) and between $\textbf{N}_{e-ci-M3}$ and $\textbf{N}_{e-ci-M4}$ (around $5^\circ$ at $\sim$07:55:33.4 UT). Firstly, the average velocity across the current sheet calculated by timing analysis of reconnecting field observed by MMS1 and MMS2 is $\sim$190.5 $kms^{-1}$. The normal direction ($\textbf{n}$) is defined as $\textbf{n}\sim\textbf{N}_{e-ci-M1}=[0.91,0.27,-0.31]_{GSE}$. Therefore, the current sheet is moving at $\textbf{V}_{SN,12}=190.5\times \textbf{N}_{e-ci-M1}=190.5\times [0.95,0.14,-0.26]_{GSE}$ $kms^{-1}$. Following the same procedure, the moving speed of the current sheet derived from MMS3 and MMS4 observations is $\textbf{V}_{SN,34}=175.5\times \textbf{N}_{e-ci-M4}=175.5\times [0.91,0.27,-0.31]_{GSE}$ $kms^{-1}$, roughly equal to $\textbf{V}_{SN,12}$. Due to only a slight difference in the moving velocity, we use $\textbf{V}_{SN,12}$ to estimate the half-thickness of the current sheet. Based on MMS1 observations, the interval of the current sheet with $|J|>\frac{1}{e^{*}}|J_{max}|$ is $\sim$0.24 s, where $e^{*}$ refers to the base of natural logarithms. Then, the half-thickness of the current sheet is estimated to be $\sim$23 km, roughly eight times the electron inertial length ($d_{e}\sim3$ $km$). The layer of bipolar $B_{L,e-ci-M1}$ between 07:55:29.57 and 07:55:30.37 UT corresponds to a half-thickness of $\sim$72.4 km, which is much larger than cold-ion gyroradius $R_{cold}\sim$8.4 km (taking $|B|\sim$106 nT and $T_{i,cold}\sim$43.8 eV) but smaller than hot-ion gyroradius $R_{hot}\sim$ 114.8 km (taking $|B|\sim$106 nT and $T_{i,hot}\sim$11589 eV). It might explain why cold ions remain magnetized, whereas hot ions decouple from the magnetic field (Figure 6h). 

We speculate that dawnward Hall $B_{Y,GSE}$ of the ion-scale reconnection reconnects with the duskward background magnetic field, resulting in the secondary reconnection. Observations only show reconnection features of electrons and cold ions. Based on the differences in characteristic length scales, there are two candidate explanations for the absence of the hot-ion response in the electron-cold ion reconnection. (1) Hot ions are not involved in the electron-cold ion reconnection. Since the half-thickness of the current sheet is smaller than $R_{hot}$, magnetic energy released by topology variations cannot effectively convert to hot-ion kinetic energy. Due to the finite Hall magnetic field in space, the length of the electron-cold ion current sheet is limited \cite{Huba2002}. Thus, there is not enough space and time for hot ions to couple to the magnetic field overall dimensions \cite{Phan2018} . (2) Hot ions are involved in the electron-cold ion reconnection but not observed. In this situation, the spacecraft cross inside the hot-ion diffusion region but outside the cold-ion diffusion region, where cold-ion jets can be detected, whereas hot-ion jets have not formed. As shown in Figure 5c, MMS3 is located far away from the X-line (at least 200 km) but does not observe hot-ion outflows. Therefore, we deduce that the first scenario, in which electron-cold ion reconnection is involved only by electrons and cold ions, is most likely.

The multiscale reconnection scenario shows different types of magnetic topologies. For the ion-scale reconnection, the magnetic shear angle is $\sim153^\circ$, suggesting an antiparallel reconnection. For the electron-cold ion reconnection, based on MMS1 observations, the ratio of the guide field to the magnetospheric reconnecting field of the current sheet $B_{M,e-ci-M1}/B_{ph}\sim5.7$ indicates a component reconnection with a large out-of-plane guide field. Therefore, the multiscale process provides evidence for the simultaneous operation of antiparallel and component reconnection.

\subsection{Other possible interpretations of the electron-cold ion current sheet}

An alternate interpretation of the electron-cold ion current sheet is a rapid transition across various boundaries that make up the magnetospheric side of the magnetopause reconnection region. A relatively thick boundary separates the electron-only boundary layer (between Figure 1\circled{1} and 1\circled{2}) from the magnetosheath ion and electron boundary layer (between Figure 1\circled{2} and 1\circled{3}). There are gradients on the magnetospheric electrons and both ions and electrons of magnetosheath origin, driving localized gradient drifts. The strong field-aligned electron jets (Figure 1c), related to a concentrated current connected to Hall currents through the ion-scale diffusion region, would drive the magnetic field variations in $X_{GSE}$ and $Y_{GSE}$ components. This interpretation is unnecessary to introduce a secondary reconnection and is more in line with the current understanding of reconnection. However, this scenario cannot explain why cold-ion jets roughly in $Y_{GSE}$ direction satisfy Walén relations in the cold-ion flow frame. More interpretations on the electron-cold ion current sheet need further study.

\section{Conclusions}

In summary, we report remarkable observations of a multiscale, multi-type dayside magnetopause reconnection observed by four MMS spacecraft on 11 September 2015. We find a new phenomenon: low-energy ions outside the diffusion region can introduce an electron-cold ion reconnecting current sheet at the edge of the ion-scale reconnection by involving a new length scale. The observations can be interpreted as a secondary reconnection dominated by electrons and cold ions (mainly in $XY_{GSE}$ plane) located at the edge of an ongoing ion-scale reconnection (mainly in $XZ_{GSE}$ plane). In contrast to ion-scale reconnection, electron and cold-ion outflows of electron-cold ion reconnection are comparable to cold-ion Alfvén speed, but hot-ion outflows are not observed. Thus, the analysis demonstrates the dominant role of cold ions and electrons in the secondary reconnection on the magnetospheric side of magnetopause without hot ions’ response. Cold ions and electrons are accelerated and heated by the secondary process. As a result, the pre-accelerated and heated cold ions and electrons may participate in the ion-scale reconnection, affecting the solar wind-magnetopause coupling. Previous studies showed cold ions decrease the reconnection rate when reaching magnetopause \cite{Walsh2013}. However, our study shows that cold ions with smaller characteristic lengths can introduce a secondary reconnection at the edge of the ion-scale magnetopause reconnection, providing a positive contribution. Therefore, the effects of cold ions on the magnetopause reconnection are underestimated and should be reconsidered.

Moreover, we find that the secondary reconnection in the presence of a strong guide field operates simultaneously with the ion-scale antiparallel reconnection. To our knowledge, reconnection models treat antiparallel and component reconnection separately and cannot address whether they operate simultaneously \cite{Fuselier2011}. It is the first observational evidence for the simultaneous coexistence of reconnection of different types. The complicated magnetic field topology could affect the reconnection rate. Further understanding of the effects of multiscale, multi-type dayside magnetopause reconnection and how often and under what circumstances the multiscale process can occur at the dayside magnetopause need more investigations in the future.

\acknowledgments

MMS data and WIND data are found online (https://spdf.gsfc.nasa.gov/). Data analysis is performed using the IRFU-Matlab analysis package available at https://github.com/irfu/irfu-matlab and the SPADAS analysis software available at http://themis.ssl.berkeley.edu. AE data are obtained from the NASA OMNIWeb (https://omniweb.gsfc.nasa.gov).

This work was supported by NSFC grants 11975038. We would like to thank the MMS spacecraft team, the WIND spacecraft team, NASA’s Coordinated Data Analysis Web (CDAWeb), the NASA OMNIWeb, and the Chinese Meridian Project for their data access. S. Q. Z. is supported by the scholarship provided by the Chinese Scholarship Council. R. Rankin acknowledges financial support from the Canadian Space Agency and NSERC. We acknowledge the plasma astrophysics group led by Huirong Yan in Deutsches Elektronen Synchrotron (DESY).


%
%

\bibliography{agusample.bib}

\begin{thebibliography}{}

\bibitem [\protect \citeauthoryear {%
Alm%
\ \protect \BOthers {.}}{%
Alm%
\ \protect \BOthers {.}}{%
{\protect \APACyear {2018}}%
}]{%
Alm2018}
\APACinsertmetastar {%
Alm2018}%
\begin{APACrefauthors}%
Alm, L.%
, Andr{\'{e}}, M.%
, Vaivads, A.%
, Khotyaintsev, Y\BPBI V.%
, Torbert, R\BPBI B.%
, Burch, J\BPBI L.%
\BDBL {}Mauk, B\BPBI H.%
\end{APACrefauthors}%
\unskip\
\newblock
\APACrefYearMonthDay{2018}{}{}.
\newblock
{\BBOQ}\APACrefatitle {{Magnetotail Hall Physics in the Presence of Cold Ions}}
  {{Magnetotail Hall Physics in the Presence of Cold Ions}}.{\BBCQ}
\newblock
\APACjournalVolNumPages{Geophysical Research Letters}{45}{20}{10,941--10,950}.
\newblock
\begin{APACrefDOI} \doi{10.1029/2018GL079857} \end{APACrefDOI}
\PrintBackRefs{\CurrentBib}

\bibitem [\protect \citeauthoryear {%
Andr{\'{e}}%
\ \protect \BOthers {.}}{%
Andr{\'{e}}%
\ \protect \BOthers {.}}{%
{\protect \APACyear {2016}}%
}]{%
Andre2016}
\APACinsertmetastar {%
Andre2016}%
\begin{APACrefauthors}%
Andr{\'{e}}, M.%
, Li, W.%
, Toledo-Redondo, S.%
, Khotyaintsev, Y\BPBI V.%
, Vaivads, A.%
, Graham, D\BPBI B.%
\BDBL {}Saito, Y.%
\end{APACrefauthors}%
\unskip\
\newblock
\APACrefYearMonthDay{2016}{}{}.
\newblock
{\BBOQ}\APACrefatitle {{Magnetic reconnection and modification of the Hall
  physics due to cold ions at the magnetopause}} {{Magnetic reconnection and
  modification of the Hall physics due to cold ions at the
  magnetopause}}.{\BBCQ}
\newblock
\APACjournalVolNumPages{Geophysical Research Letters}{43}{13}{6705--6712}.
\newblock
\begin{APACrefDOI} \doi{10.1002/2016GL069665} \end{APACrefDOI}
\PrintBackRefs{\CurrentBib}

\bibitem [\protect \citeauthoryear {%
Angelopoulos%
\ \protect \BOthers {.}}{%
Angelopoulos%
\ \protect \BOthers {.}}{%
{\protect \APACyear {2008}}%
}]{%
Angelopoulos2008}
\APACinsertmetastar {%
Angelopoulos2008}%
\begin{APACrefauthors}%
Angelopoulos, V.%
, McFadden, J\BPBI P.%
, Larson, D.%
, Carlson, C\BPBI W.%
, Mende, S\BPBI B.%
, Frey, H.%
\BDBL {}Kepko, L.%
\end{APACrefauthors}%
\unskip\
\newblock
\APACrefYearMonthDay{2008}{aug}{}.
\newblock
{\BBOQ}\APACrefatitle {{Tail Reconnection Triggering Substorm Onset}} {{Tail
  Reconnection Triggering Substorm Onset}}.{\BBCQ}
\newblock
\APACjournalVolNumPages{Science}{321}{5891}{931--935}.
\newblock
\begin{APACrefURL}
  \url{https://www.sciencemag.org/lookup/doi/10.1126/science.1160495}
  \end{APACrefURL}
\newblock
\begin{APACrefDOI} \doi{10.1126/science.1160495} \end{APACrefDOI}
\PrintBackRefs{\CurrentBib}

\bibitem [\protect \citeauthoryear {%
Borovsky%
\ \BBA {} Denton%
}{%
Borovsky%
\ \BBA {} Denton%
}{%
{\protect \APACyear {2008}}%
}]{%
Borovsky2008}
\APACinsertmetastar {%
Borovsky2008}%
\begin{APACrefauthors}%
Borovsky, J\BPBI E.%
\BCBT {}\ \BBA {} Denton, M\BPBI H.%
\end{APACrefauthors}%
\unskip\
\newblock
\APACrefYearMonthDay{2008}{}{}.
\newblock
{\BBOQ}\APACrefatitle {{A statistical look at plasmaspheric drainage plumes}}
  {{A statistical look at plasmaspheric drainage plumes}}.{\BBCQ}
\newblock
\APACjournalVolNumPages{Journal of Geophysical Research: Space
  Physics}{113}{9}{1--23}.
\newblock
\begin{APACrefDOI} \doi{10.1029/2007JA012994} \end{APACrefDOI}
\PrintBackRefs{\CurrentBib}

\bibitem [\protect \citeauthoryear {%
Burch%
, Moore%
, Torbert%
\BCBL {}\ \BBA {} Giles%
}{%
Burch%
, Moore%
\BCBL {}\ \protect \BOthers {.}}{%
{\protect \APACyear {2016}}%
}]{%
Burch2016a}
\APACinsertmetastar {%
Burch2016a}%
\begin{APACrefauthors}%
Burch, J\BPBI L.%
, Moore, T\BPBI E.%
, Torbert, R\BPBI B.%
\BCBL {}\ \BBA {} Giles, B\BPBI L.%
\end{APACrefauthors}%
\unskip\
\newblock
\APACrefYearMonthDay{2016}{}{}.
\newblock
{\BBOQ}\APACrefatitle {{Magnetospheric Multiscale Overview and Science
  Objectives}} {{Magnetospheric Multiscale Overview and Science
  Objectives}}.{\BBCQ}
\newblock
\APACjournalVolNumPages{Space Science Reviews}{199}{1-4}{5--21}.
\newblock
\begin{APACrefURL} \url{http://dx.doi.org/10.1007/s11214-015-0164-9}
  \end{APACrefURL}
\newblock
\begin{APACrefDOI} \doi{10.1007/s11214-015-0164-9} \end{APACrefDOI}
\PrintBackRefs{\CurrentBib}

\bibitem [\protect \citeauthoryear {%
Burch%
, Torbert%
\BCBL {}\ \protect \BOthers {.}}{%
Burch%
, Torbert%
\BCBL {}\ \protect \BOthers {.}}{%
{\protect \APACyear {2016}}%
}]{%
Burch2016b}
\APACinsertmetastar {%
Burch2016b}%
\begin{APACrefauthors}%
Burch, J\BPBI L.%
, Torbert, R\BPBI B.%
, Phan, T\BPBI D.%
, Chen, L\BPBI J.%
, Moore, T\BPBI E.%
, Ergun, R\BPBI E.%
\BDBL {}Chandler, M.%
\end{APACrefauthors}%
\unskip\
\newblock
\APACrefYearMonthDay{2016}{}{}.
\newblock
{\BBOQ}\APACrefatitle {{Electron-scale measurements of magnetic reconnection in
  space}} {{Electron-scale measurements of magnetic reconnection in
  space}}.{\BBCQ}
\newblock
\APACjournalVolNumPages{Science}{352}{6290}{}.
\newblock
\begin{APACrefDOI} \doi{10.1126/science.aaf2939} \end{APACrefDOI}
\PrintBackRefs{\CurrentBib}

\bibitem [\protect \citeauthoryear {%
Cassak%
\ \BBA {} Shay%
}{%
Cassak%
\ \BBA {} Shay%
}{%
{\protect \APACyear {2007}}%
}]{%
Cassak2007}
\APACinsertmetastar {%
Cassak2007}%
\begin{APACrefauthors}%
Cassak, P\BPBI A.%
\BCBT {}\ \BBA {} Shay, M\BPBI A.%
\end{APACrefauthors}%
\unskip\
\newblock
\APACrefYearMonthDay{2007}{}{}.
\newblock
{\BBOQ}\APACrefatitle {{Scaling of asymmetric magnetic reconnection: General
  theory and collisional simulations}} {{Scaling of asymmetric magnetic
  reconnection: General theory and collisional simulations}}.{\BBCQ}
\newblock
\APACjournalVolNumPages{Physics of Plasmas}{14}{10}{}.
\newblock
\begin{APACrefDOI} \doi{10.1063/1.2795630} \end{APACrefDOI}
\PrintBackRefs{\CurrentBib}

\bibitem [\protect \citeauthoryear {%
Chappell%
, Huddleston%
, Moore%
, Giles%
\BCBL {}\ \BBA {} Delcourt%
}{%
Chappell%
\ \protect \BOthers {.}}{%
{\protect \APACyear {2008}}%
}]{%
Chappell2008}
\APACinsertmetastar {%
Chappell2008}%
\begin{APACrefauthors}%
Chappell, C\BPBI R.%
, Huddleston, M\BPBI M.%
, Moore, T\BPBI E.%
, Giles, B\BPBI L.%
\BCBL {}\ \BBA {} Delcourt, D\BPBI C.%
\end{APACrefauthors}%
\unskip\
\newblock
\APACrefYearMonthDay{2008}{}{}.
\newblock
{\BBOQ}\APACrefatitle {{Observations of the warm plasma cloak and an
  explanation of its formation in the magnetosphere}} {{Observations of the
  warm plasma cloak and an explanation of its formation in the
  magnetosphere}}.{\BBCQ}
\newblock
\APACjournalVolNumPages{Journal of Geophysical Research: Space
  Physics}{113}{9}{1--21}.
\newblock
\begin{APACrefDOI} \doi{10.1029/2007JA012945} \end{APACrefDOI}
\PrintBackRefs{\CurrentBib}

\bibitem [\protect \citeauthoryear {%
L\BPBI J.~Chen%
, Hesse%
, Wang%
, Bessho%
\BCBL {}\ \BBA {} Daughton%
}{%
L\BPBI J.~Chen%
\ \protect \BOthers {.}}{%
{\protect \APACyear {2016}}%
}]{%
Chen2016}
\APACinsertmetastar {%
Chen2016}%
\begin{APACrefauthors}%
Chen, L\BPBI J.%
, Hesse, M.%
, Wang, S.%
, Bessho, N.%
\BCBL {}\ \BBA {} Daughton, W.%
\end{APACrefauthors}%
\unskip\
\newblock
\APACrefYearMonthDay{2016}{}{}.
\newblock
{\BBOQ}\APACrefatitle {{Electron energization and structure of the diffusion
  region during asymmetric reconnection}} {{Electron energization and structure
  of the diffusion region during asymmetric reconnection}}.{\BBCQ}
\newblock
\APACjournalVolNumPages{Geophysical Research Letters}{43}{6}{2405--2412}.
\newblock
\begin{APACrefDOI} \doi{10.1002/2016GL068243} \end{APACrefDOI}
\PrintBackRefs{\CurrentBib}

\bibitem [\protect \citeauthoryear {%
S\BPBI H.~Chen%
\ \BBA {} Moore%
}{%
S\BPBI H.~Chen%
\ \BBA {} Moore%
}{%
{\protect \APACyear {2004}}%
}]{%
Chen2004}
\APACinsertmetastar {%
Chen2004}%
\begin{APACrefauthors}%
Chen, S\BPBI H.%
\BCBT {}\ \BBA {} Moore, T\BPBI E.%
\end{APACrefauthors}%
\unskip\
\newblock
\APACrefYearMonthDay{2004}{}{}.
\newblock
{\BBOQ}\APACrefatitle {{Dayside flow bursts in the Earth's outer
  magnetosphere}} {{Dayside flow bursts in the Earth's outer
  magnetosphere}}.{\BBCQ}
\newblock
\APACjournalVolNumPages{Journal of Geophysical Research: Space
  Physics}{109}{A3}{1--15}.
\newblock
\begin{APACrefDOI} \doi{10.1029/2003JA010007} \end{APACrefDOI}
\PrintBackRefs{\CurrentBib}

\bibitem [\protect \citeauthoryear {%
Darrouzet%
, {De Keyser}%
, Dacrau%
, {El Lemdani-Mazouz}%
\BCBL {}\ \BBA {} Vallires%
}{%
Darrouzet%
\ \protect \BOthers {.}}{%
{\protect \APACyear {2008}}%
}]{%
Darrouzet2008}
\APACinsertmetastar {%
Darrouzet2008}%
\begin{APACrefauthors}%
Darrouzet, F.%
, {De Keyser}, J.%
, Dacrau, P\BPBI M.%
, {El Lemdani-Mazouz}, F.%
\BCBL {}\ \BBA {} Vallires, X.%
\end{APACrefauthors}%
\unskip\
\newblock
\APACrefYearMonthDay{2008}{}{}.
\newblock
{\BBOQ}\APACrefatitle {{Statistical analysis of plasmaspheric plumes with
  Cluster/WHISPER observations}} {{Statistical analysis of plasmaspheric plumes
  with Cluster/WHISPER observations}}.{\BBCQ}
\newblock
\APACjournalVolNumPages{Annales Geophysicae}{26}{8}{2403--2417}.
\newblock
\begin{APACrefDOI} \doi{10.5194/angeo-26-2403-2008} \end{APACrefDOI}
\PrintBackRefs{\CurrentBib}

\bibitem [\protect \citeauthoryear {%
Dungey%
}{%
Dungey%
}{%
{\protect \APACyear {1961}}%
}]{%
Dungey1961}
\APACinsertmetastar {%
Dungey1961}%
\begin{APACrefauthors}%
Dungey, J\BPBI W.%
\end{APACrefauthors}%
\unskip\
\newblock
\APACrefYearMonthDay{1961}{}{}.
\newblock
{\BBOQ}\APACrefatitle {{Interplanetary magnetic field and the auroral zones}}
  {{Interplanetary magnetic field and the auroral zones}}.{\BBCQ}
\newblock
\APACjournalVolNumPages{Physical Review Letters}{6}{2}{47--48}.
\newblock
\begin{APACrefDOI} \doi{10.1103/PhysRevLett.6.47} \end{APACrefDOI}
\PrintBackRefs{\CurrentBib}

\bibitem [\protect \citeauthoryear {%
Engwall%
\ \protect \BOthers {.}}{%
Engwall%
\ \protect \BOthers {.}}{%
{\protect \APACyear {2009}}%
}]{%
Engwall2009}
\APACinsertmetastar {%
Engwall2009}%
\begin{APACrefauthors}%
Engwall, E.%
, Eriksson, A\BPBI I.%
, Cully, C\BPBI M.%
, Andr{\'{e}}, M.%
, Torbert, R.%
\BCBL {}\ \BBA {} Vaith, H.%
\end{APACrefauthors}%
\unskip\
\newblock
\APACrefYearMonthDay{2009}{}{}.
\newblock
{\BBOQ}\APACrefatitle {{Earth's ionospheric outflow dominated by hidden cold
  plasma}} {{Earth's ionospheric outflow dominated by hidden cold
  plasma}}.{\BBCQ}
\newblock
\APACjournalVolNumPages{Nature Geoscience}{2}{1}{24--27}.
\newblock
\begin{APACrefURL} \url{http://dx.doi.org/10.1038/ngeo387} \end{APACrefURL}
\newblock
\begin{APACrefDOI} \doi{10.1038/ngeo387} \end{APACrefDOI}
\PrintBackRefs{\CurrentBib}

\bibitem [\protect \citeauthoryear {%
Ergun%
\ \protect \BOthers {.}}{%
Ergun%
\ \protect \BOthers {.}}{%
{\protect \APACyear {2016}}%
}]{%
Ergun2016}
\APACinsertmetastar {%
Ergun2016}%
\begin{APACrefauthors}%
Ergun, R\BPBI E.%
, Tucker, S.%
, Westfall, J.%
, Goodrich, K\BPBI A.%
, Malaspina, D\BPBI M.%
, Summers, D.%
\BDBL {}Cully, C\BPBI M.%
\end{APACrefauthors}%
\unskip\
\newblock
\APACrefYearMonthDay{2016}{}{}.
\newblock
{\BBOQ}\APACrefatitle {{The Axial Double Probe and Fields Signal Processing for
  the MMS Mission}} {{The Axial Double Probe and Fields Signal Processing for
  the MMS Mission}}.{\BBCQ}
\newblock
\APACjournalVolNumPages{Space Science Reviews}{199}{1-4}{167--188}.
\newblock
\begin{APACrefURL} \url{http://dx.doi.org/10.1007/s11214-014-0115-x}
  \end{APACrefURL}
\newblock
\begin{APACrefDOI} \doi{10.1007/s11214-014-0115-x} \end{APACrefDOI}
\PrintBackRefs{\CurrentBib}

\bibitem [\protect \citeauthoryear {%
Fuselier%
\ \BBA {} Lewis%
}{%
Fuselier%
\ \BBA {} Lewis%
}{%
{\protect \APACyear {2011}}%
}]{%
Fuselier2011}
\APACinsertmetastar {%
Fuselier2011}%
\begin{APACrefauthors}%
Fuselier, S\BPBI A.%
\BCBT {}\ \BBA {} Lewis, W\BPBI S.%
\end{APACrefauthors}%
\unskip\
\newblock
\APACrefYearMonthDay{2011}{}{}.
\newblock
{\BBOQ}\APACrefatitle {{Properties of near-earth magnetic reconnection from
  in-situ observations}} {{Properties of near-earth magnetic reconnection from
  in-situ observations}}.{\BBCQ}
\newblock
\APACjournalVolNumPages{Space Science Reviews}{160}{1-4}{95--121}.
\newblock
\begin{APACrefDOI} \doi{10.1007/s11214-011-9820-x} \end{APACrefDOI}
\PrintBackRefs{\CurrentBib}

\bibitem [\protect \citeauthoryear {%
Fuselier%
\ \protect \BOthers {.}}{%
Fuselier%
\ \protect \BOthers {.}}{%
{\protect \APACyear {2016}}%
}]{%
Fuselier2016}
\APACinsertmetastar {%
Fuselier2016}%
\begin{APACrefauthors}%
Fuselier, S\BPBI A.%
, Lewis, W\BPBI S.%
, Schiff, C.%
, Ergun, R.%
, Burch, J\BPBI L.%
, Petrinec, S\BPBI M.%
\BCBL {}\ \BBA {} Trattner, K\BPBI J.%
\end{APACrefauthors}%
\unskip\
\newblock
\APACrefYearMonthDay{2016}{}{}.
\newblock
{\BBOQ}\APACrefatitle {{Magnetospheric Multiscale Science Mission Profile and
  Operations}} {{Magnetospheric Multiscale Science Mission Profile and
  Operations}}.{\BBCQ}
\newblock
\APACjournalVolNumPages{Space Science Reviews}{199}{1-4}{77--103}.
\newblock
\begin{APACrefURL} \url{http://dx.doi.org/10.1007/s11214-014-0087-x}
  \end{APACrefURL}
\newblock
\begin{APACrefDOI} \doi{10.1007/s11214-014-0087-x} \end{APACrefDOI}
\PrintBackRefs{\CurrentBib}

\bibitem [\protect \citeauthoryear {%
Goldman%
, Newman%
\BCBL {}\ \BBA {} Pritchett%
}{%
Goldman%
\ \protect \BOthers {.}}{%
{\protect \APACyear {2008}}%
}]{%
Goldman2008}
\APACinsertmetastar {%
Goldman2008}%
\begin{APACrefauthors}%
Goldman, M\BPBI V.%
, Newman, D\BPBI L.%
\BCBL {}\ \BBA {} Pritchett, P.%
\end{APACrefauthors}%
\unskip\
\newblock
\APACrefYearMonthDay{2008}{nov}{}.
\newblock
{\BBOQ}\APACrefatitle {{Vlasov simulations of electron holes driven by particle
  distributions from PIC reconnection simulations with a guide field}} {{Vlasov
  simulations of electron holes driven by particle distributions from PIC
  reconnection simulations with a guide field}}.{\BBCQ}
\newblock
\APACjournalVolNumPages{Geophysical Research Letters}{35}{22}{L22109}.
\newblock
\begin{APACrefURL} \url{http://doi.wiley.com/10.1029/2008GL035608}
  \end{APACrefURL}
\newblock
\begin{APACrefDOI} \doi{10.1029/2008GL035608} \end{APACrefDOI}
\PrintBackRefs{\CurrentBib}

\bibitem [\protect \citeauthoryear {%
Gosling%
\ \BBA {} Phan%
}{%
Gosling%
\ \BBA {} Phan%
}{%
{\protect \APACyear {2013}}%
}]{%
Gosling2013}
\APACinsertmetastar {%
Gosling2013}%
\begin{APACrefauthors}%
Gosling, J\BPBI T.%
\BCBT {}\ \BBA {} Phan, T\BPBI D.%
\end{APACrefauthors}%
\unskip\
\newblock
\APACrefYearMonthDay{2013}{}{}.
\newblock
{\BBOQ}\APACrefatitle {{Magnetic reconnection in the solar wind at current
  sheets associated with extremely small field shear angles}} {{Magnetic
  reconnection in the solar wind at current sheets associated with extremely
  small field shear angles}}.{\BBCQ}
\newblock
\APACjournalVolNumPages{Astrophysical Journal Letters}{763}{2}{1987--1990}.
\newblock
\begin{APACrefDOI} \doi{10.1088/2041-8205/763/2/L39} \end{APACrefDOI}
\PrintBackRefs{\CurrentBib}

\bibitem [\protect \citeauthoryear {%
Huba%
\ \BBA {} Rudakov%
}{%
Huba%
\ \BBA {} Rudakov%
}{%
{\protect \APACyear {2002}}%
}]{%
Huba2002}
\APACinsertmetastar {%
Huba2002}%
\begin{APACrefauthors}%
Huba, J\BPBI D.%
\BCBT {}\ \BBA {} Rudakov, L\BPBI I.%
\end{APACrefauthors}%
\unskip\
\newblock
\APACrefYearMonthDay{2002}{}{}.
\newblock
{\BBOQ}\APACrefatitle {{Three-dimensional hall magnetic reconnection}}
  {{Three-dimensional hall magnetic reconnection}}.{\BBCQ}
\newblock
\APACjournalVolNumPages{Physics of Plasmas}{9}{11}{4435}.
\newblock
\begin{APACrefDOI} \doi{10.1063/1.1514970} \end{APACrefDOI}
\PrintBackRefs{\CurrentBib}

\bibitem [\protect \citeauthoryear {%
Hwang%
\ \protect \BOthers {.}}{%
Hwang%
\ \protect \BOthers {.}}{%
{\protect \APACyear {2017}}%
}]{%
Hwang2017}
\APACinsertmetastar {%
Hwang2017}%
\begin{APACrefauthors}%
Hwang, K\BPBI J.%
, Sibeck, D\BPBI G.%
, Choi, E.%
, Chen, L\BPBI J.%
, Ergun, R\BPBI E.%
, Khotyaintsev, Y.%
\BDBL {}Torbert, R\BPBI B.%
\end{APACrefauthors}%
\unskip\
\newblock
\APACrefYearMonthDay{2017}{}{}.
\newblock
{\BBOQ}\APACrefatitle {{Magnetospheric Multiscale mission observations of the
  outer electron diffusion region}} {{Magnetospheric Multiscale mission
  observations of the outer electron diffusion region}}.{\BBCQ}
\newblock
\APACjournalVolNumPages{Geophysical Research Letters}{44}{5}{2049--2059}.
\newblock
\begin{APACrefDOI} \doi{10.1002/2017GL072830} \end{APACrefDOI}
\PrintBackRefs{\CurrentBib}

\bibitem [\protect \citeauthoryear {%
Lapenta%
, Markidis%
, Divin%
, Goldman%
\BCBL {}\ \BBA {} Newman%
}{%
Lapenta%
\ \protect \BOthers {.}}{%
{\protect \APACyear {2011}}%
}]{%
Lapenta2011}
\APACinsertmetastar {%
Lapenta2011}%
\begin{APACrefauthors}%
Lapenta, G.%
, Markidis, S.%
, Divin, A.%
, Goldman, M\BPBI V.%
\BCBL {}\ \BBA {} Newman, D\BPBI L.%
\end{APACrefauthors}%
\unskip\
\newblock
\APACrefYearMonthDay{2011}{sep}{}.
\newblock
{\BBOQ}\APACrefatitle {{Bipolar electric field signatures of reconnection
  separatrices for a hydrogen plasma at realistic guide fields}} {{Bipolar
  electric field signatures of reconnection separatrices for a hydrogen plasma
  at realistic guide fields}}.{\BBCQ}
\newblock
\APACjournalVolNumPages{Geophysical Research Letters}{38}{17}{n/a--n/a}.
\newblock
\begin{APACrefURL} \url{http://doi.wiley.com/10.1029/2011GL048572}
  \end{APACrefURL}
\newblock
\begin{APACrefDOI} \doi{10.1029/2011GL048572} \end{APACrefDOI}
\PrintBackRefs{\CurrentBib}

\bibitem [\protect \citeauthoryear {%
Lee%
\ \protect \BOthers {.}}{%
Lee%
\ \protect \BOthers {.}}{%
{\protect \APACyear {2016}}%
}]{%
Lee2016}
\APACinsertmetastar {%
Lee2016}%
\begin{APACrefauthors}%
Lee, S\BPBI H.%
, Zhang, H.%
, Zong, Q\BPBI G.%
, Otto, A.%
, R{\`{e}}me, H.%
\BCBL {}\ \BBA {} Liebert, E.%
\end{APACrefauthors}%
\unskip\
\newblock
\APACrefYearMonthDay{2016}{}{}.
\newblock
{\BBOQ}\APACrefatitle {{A statistical study of plasmaspheric plumes and
  ionospheric outflows observed at the dayside magnetopause}} {{A statistical
  study of plasmaspheric plumes and ionospheric outflows observed at the
  dayside magnetopause}}.{\BBCQ}
\newblock
\APACjournalVolNumPages{Journal of Geophysical Research A: Space
  Physics}{121}{1}{492--506}.
\newblock
\begin{APACrefDOI} \doi{10.1002/2015JA021540} \end{APACrefDOI}
\PrintBackRefs{\CurrentBib}

\bibitem [\protect \citeauthoryear {%
Lee%
\ \protect \BOthers {.}}{%
Lee%
\ \protect \BOthers {.}}{%
{\protect \APACyear {2014}}%
}]{%
Zhang2014}
\APACinsertmetastar {%
Zhang2014}%
\begin{APACrefauthors}%
Lee, S\BPBI H.%
, Zhang, H.%
, Zong, Q\BHBI G.%
, Otto, A.%
, Sibeck, D\BPBI G.%
, Wang, Y.%
\BDBL {}R{\`{e}}me, H.%
\end{APACrefauthors}%
\unskip\
\newblock
\APACrefYearMonthDay{2014}{mar}{}.
\newblock
{\BBOQ}\APACrefatitle {{Plasma and energetic particle behaviors during
  asymmetric magnetic reconnection at the magnetopause}} {{Plasma and energetic
  particle behaviors during asymmetric magnetic reconnection at the
  magnetopause}}.{\BBCQ}
\newblock
\APACjournalVolNumPages{Journal of Geophysical Research: Space
  Physics}{119}{3}{1658--1672}.
\newblock
\begin{APACrefURL} \url{http://doi.wiley.com/10.1002/2013JA019168}
  \end{APACrefURL}
\newblock
\begin{APACrefDOI} \doi{10.1002/2013JA019168} \end{APACrefDOI}
\PrintBackRefs{\CurrentBib}

\bibitem [\protect \citeauthoryear {%
Lepping%
\ \protect \BOthers {.}}{%
Lepping%
\ \protect \BOthers {.}}{%
{\protect \APACyear {1995}}%
}]{%
Lepping1995}
\APACinsertmetastar {%
Lepping1995}%
\begin{APACrefauthors}%
Lepping, R\BPBI P.%
, Acũna, M\BPBI H.%
, Burlaga, L\BPBI F.%
, Farrell, W\BPBI M.%
, Slavin, J\BPBI A.%
, Schatten, K\BPBI H.%
\BDBL {}Worley, E\BPBI M.%
\end{APACrefauthors}%
\unskip\
\newblock
\APACrefYearMonthDay{1995}{feb}{}.
\newblock
{\BBOQ}\APACrefatitle {{The WIND magnetic field investigation}} {{The WIND
  magnetic field investigation}}.{\BBCQ}
\newblock
\APACjournalVolNumPages{Space Science Reviews}{71}{1-4}{207--229}.
\newblock
\begin{APACrefDOI} \doi{10.1007/BF00751330} \end{APACrefDOI}
\PrintBackRefs{\CurrentBib}

\bibitem [\protect \citeauthoryear {%
Liang%
\ \protect \BOthers {.}}{%
Liang%
\ \protect \BOthers {.}}{%
{\protect \APACyear {2015}}%
}]{%
Liang2015}
\APACinsertmetastar {%
Liang2015}%
\begin{APACrefauthors}%
Liang, J.%
, Donovan, E.%
, Nishimura, Y.%
, Yang, B.%
, Spanswick, E.%
, Asamura, K.%
\BDBL {}Redmon, R.%
\end{APACrefauthors}%
\unskip\
\newblock
\APACrefYearMonthDay{2015}{jul}{}.
\newblock
{\BBOQ}\APACrefatitle {{Low-energy ion precipitation structures associated with
  pulsating auroral patches}} {{Low-energy ion precipitation structures
  associated with pulsating auroral patches}}.{\BBCQ}
\newblock
\APACjournalVolNumPages{Journal of Geophysical Research: Space
  Physics}{120}{7}{5408--5431}.
\newblock
\begin{APACrefURL} \url{http://doi.wiley.com/10.1002/2015JA021094}
  \end{APACrefURL}
\newblock
\begin{APACrefDOI} \doi{10.1002/2015JA021094} \end{APACrefDOI}
\PrintBackRefs{\CurrentBib}

\bibitem [\protect \citeauthoryear {%
Lin%
\ \protect \BOthers {.}}{%
Lin%
\ \protect \BOthers {.}}{%
{\protect \APACyear {1995}}%
}]{%
Lin1995}
\APACinsertmetastar {%
Lin1995}%
\begin{APACrefauthors}%
Lin, R\BPBI P.%
, Anderson, K\BPBI A.%
, Ashford, S.%
, Carlson, C.%
, Curtis, D.%
, Ergun, R.%
\BDBL {}Paschmann, G.%
\end{APACrefauthors}%
\unskip\
\newblock
\APACrefYearMonthDay{1995}{}{}.
\newblock
{\BBOQ}\APACrefatitle {{A three-dimensional plasma and energetic particle
  investigation for the wind spacecraft}} {{A three-dimensional plasma and
  energetic particle investigation for the wind spacecraft}}.{\BBCQ}
\newblock
\APACjournalVolNumPages{Space Science Reviews}{71}{1-4}{125--153}.
\newblock
\begin{APACrefDOI} \doi{10.1007/BF00751328} \end{APACrefDOI}
\PrintBackRefs{\CurrentBib}

\bibitem [\protect \citeauthoryear {%
Lindqvist%
\ \protect \BOthers {.}}{%
Lindqvist%
\ \protect \BOthers {.}}{%
{\protect \APACyear {2016}}%
}]{%
Lindqvist2016}
\APACinsertmetastar {%
Lindqvist2016}%
\begin{APACrefauthors}%
Lindqvist, P\BPBI A.%
, Olsson, G.%
, Torbert, R\BPBI B.%
, King, B.%
, Granoff, M.%
, Rau, D.%
\BDBL {}Tucker, S.%
\end{APACrefauthors}%
\unskip\
\newblock
\APACrefYearMonthDay{2016}{}{}.
\newblock
{\BBOQ}\APACrefatitle {{The Spin-Plane Double Probe Electric Field Instrument
  for MMS}} {{The Spin-Plane Double Probe Electric Field Instrument for
  MMS}}.{\BBCQ}
\newblock
\APACjournalVolNumPages{Space Science Reviews}{199}{1-4}{137--165}.
\newblock
\begin{APACrefURL} \url{http://dx.doi.org/10.1007/s11214-014-0116-9}
  \end{APACrefURL}
\newblock
\begin{APACrefDOI} \doi{10.1007/s11214-014-0116-9} \end{APACrefDOI}
\PrintBackRefs{\CurrentBib}

\bibitem [\protect \citeauthoryear {%
Moore%
, Fok%
, Delcourt%
, Slinker%
\BCBL {}\ \BBA {} Fedder%
}{%
Moore%
\ \protect \BOthers {.}}{%
{\protect \APACyear {2008}}%
}]{%
Moore2008}
\APACinsertmetastar {%
Moore2008}%
\begin{APACrefauthors}%
Moore, T\BPBI E.%
, Fok, M\BPBI C.%
, Delcourt, D\BPBI C.%
, Slinker, S\BPBI P.%
\BCBL {}\ \BBA {} Fedder, J\BPBI A.%
\end{APACrefauthors}%
\unskip\
\newblock
\APACrefYearMonthDay{2008}{}{}.
\newblock
{\BBOQ}\APACrefatitle {{Plasma plume circulation and impact in an MHD
  substorm}} {{Plasma plume circulation and impact in an MHD substorm}}.{\BBCQ}
\newblock
\APACjournalVolNumPages{Journal of Geophysical Research: Space
  Physics}{113}{6}{1--9}.
\newblock
\begin{APACrefDOI} \doi{10.1029/2008JA013050} \end{APACrefDOI}
\PrintBackRefs{\CurrentBib}

\bibitem [\protect \citeauthoryear {%
Paschmann%
\ \BBA {} Sonnerup%
}{%
Paschmann%
\ \BBA {} Sonnerup%
}{%
{\protect \APACyear {2000}}%
}]{%
Paschmann2000}
\APACinsertmetastar {%
Paschmann2000}%
\begin{APACrefauthors}%
Paschmann, G.%
\BCBT {}\ \BBA {} Sonnerup, B\BPBI U\BPBI {\"{O}}.%
\end{APACrefauthors}%
\unskip\
\newblock
\APACrefYearMonthDay{2000}{}{}.
\newblock
{\BBOQ}\APACrefatitle {{Proper Frame Determination and Walen Test}} {{Proper
  Frame Determination and Walen Test}}.{\BBCQ}
\newblock
\APACjournalVolNumPages{}{}{}{65--74}.
\newblock
\begin{APACrefDOI} \doi{http://www.issibern.ch/forads/sr-008-07.pdf}
  \end{APACrefDOI}
\PrintBackRefs{\CurrentBib}

\bibitem [\protect \citeauthoryear {%
Phan%
\ \protect \BOthers {.}}{%
Phan%
\ \protect \BOthers {.}}{%
{\protect \APACyear {2018}}%
}]{%
Phan2018}
\APACinsertmetastar {%
Phan2018}%
\begin{APACrefauthors}%
Phan, T\BPBI D.%
, Eastwood, J\BPBI P.%
, Shay, M\BPBI A.%
, Drake, J\BPBI F.%
, Sonnerup, B\BPBI U.%
, Fujimoto, M.%
\BDBL {}Magnes, W.%
\end{APACrefauthors}%
\unskip\
\newblock
\APACrefYearMonthDay{2018}{}{}.
\newblock
{\BBOQ}\APACrefatitle {{Electron magnetic reconnection without ion coupling in
  Earth's turbulent magnetosheath}} {{Electron magnetic reconnection without
  ion coupling in Earth's turbulent magnetosheath}}.{\BBCQ}
\newblock
\APACjournalVolNumPages{Nature}{557}{7704}{202--206}.
\newblock
\begin{APACrefDOI} \doi{10.1038/s41586-018-0091-5} \end{APACrefDOI}
\PrintBackRefs{\CurrentBib}

\bibitem [\protect \citeauthoryear {%
Pollock%
\ \protect \BOthers {.}}{%
Pollock%
\ \protect \BOthers {.}}{%
{\protect \APACyear {2016}}%
}]{%
Pollock2016}
\APACinsertmetastar {%
Pollock2016}%
\begin{APACrefauthors}%
Pollock, C.%
, Moore, T.%
, Jacques, A.%
, Burch, J.%
, Gliese, U.%
, Saito, Y.%
\BDBL {}Zeuch, M.%
\end{APACrefauthors}%
\unskip\
\newblock
\APACrefYearMonthDay{2016}{}{}.
\newblock
{\BBOQ}\APACrefatitle {{Fast Plasma Investigation for Magnetospheric
  Multiscale}} {{Fast Plasma Investigation for Magnetospheric
  Multiscale}}.{\BBCQ}
\newblock
\APACjournalVolNumPages{Space Science Reviews}{199}{1-4}{331--406}.
\newblock
\begin{APACrefURL} \url{http://dx.doi.org/10.1007/s11214-016-0245-4}
  \end{APACrefURL}
\newblock
\begin{APACrefDOI} \doi{10.1007/s11214-016-0245-4} \end{APACrefDOI}
\PrintBackRefs{\CurrentBib}

\bibitem [\protect \citeauthoryear {%
Russell%
\ \protect \BOthers {.}}{%
Russell%
\ \protect \BOthers {.}}{%
{\protect \APACyear {2016}}%
}]{%
Russell2016}
\APACinsertmetastar {%
Russell2016}%
\begin{APACrefauthors}%
Russell, C\BPBI T.%
, Anderson, B\BPBI J.%
, Baumjohann, W.%
, Bromund, K\BPBI R.%
, Dearborn, D.%
, Fischer, D.%
\BDBL {}Richter, I.%
\end{APACrefauthors}%
\unskip\
\newblock
\APACrefYearMonthDay{2016}{}{}.
\newblock
{\BBOQ}\APACrefatitle {{The Magnetospheric Multiscale Magnetometers}} {{The
  Magnetospheric Multiscale Magnetometers}}.{\BBCQ}
\newblock
\APACjournalVolNumPages{Space Science Reviews}{199}{1-4}{189--256}.
\newblock
\begin{APACrefDOI} \doi{10.1007/s11214-014-0057-3} \end{APACrefDOI}
\PrintBackRefs{\CurrentBib}

\bibitem [\protect \citeauthoryear {%
B.~Sonnerup%
\ \BBA {} Scheible%
}{%
B.~Sonnerup%
\ \BBA {} Scheible%
}{%
{\protect \APACyear {1998}}%
}]{%
Sonnerup1998}
\APACinsertmetastar {%
Sonnerup1998}%
\begin{APACrefauthors}%
Sonnerup, B.%
\BCBT {}\ \BBA {} Scheible, M.%
\end{APACrefauthors}%
\unskip\
\newblock
\APACrefYearMonthDay{1998}{}{}.
\newblock
{\BBOQ}\APACrefatitle {{Minimum and maximum variance analysis}} {{Minimum and
  maximum variance analysis}}.{\BBCQ}
\newblock
\APACjournalVolNumPages{Analysis methods for multi-spacecraft
  data}{001}{}{185--220}.
\newblock
\begin{APACrefURL} \url{http://www.issibern.ch/forads/sr-001-08.pdf}
  \end{APACrefURL}
\PrintBackRefs{\CurrentBib}

\bibitem [\protect \citeauthoryear {%
B\BPBI U\BPBI {\"{O}}.~Sonnerup%
, Papamastorakis%
, Paschmann%
\BCBL {}\ \BBA {} L{\"{u}}hr%
}{%
B\BPBI U\BPBI {\"{O}}.~Sonnerup%
\ \protect \BOthers {.}}{%
{\protect \APACyear {1987}}%
}]{%
Sonnerup1987}
\APACinsertmetastar {%
Sonnerup1987}%
\begin{APACrefauthors}%
Sonnerup, B\BPBI U\BPBI {\"{O}}.%
, Papamastorakis, I.%
, Paschmann, G.%
\BCBL {}\ \BBA {} L{\"{u}}hr, H.%
\end{APACrefauthors}%
\unskip\
\newblock
\APACrefYearMonthDay{1987}{}{}.
\newblock
{\BBOQ}\APACrefatitle {{Magnetopause properties from AMPTE/IRM observations of
  the convection electric field: Method development}} {{Magnetopause properties
  from AMPTE/IRM observations of the convection electric field: Method
  development}}.{\BBCQ}
\newblock
\APACjournalVolNumPages{Journal of Geophysical Research}{92}{A11}{12137}.
\newblock
\begin{APACrefDOI} \doi{10.1029/ja092ia11p12137} \end{APACrefDOI}
\PrintBackRefs{\CurrentBib}

\bibitem [\protect \citeauthoryear {%
Tanaka%
\ \protect \BOthers {.}}{%
Tanaka%
\ \protect \BOthers {.}}{%
{\protect \APACyear {2008}}%
}]{%
Tanaka2008}
\APACinsertmetastar {%
Tanaka2008}%
\begin{APACrefauthors}%
Tanaka, K\BPBI G.%
, Retin{\`{o}}, A.%
, Asano, Y.%
, Fujimoto, M.%
, Shinohara, I.%
, Vaivads, A.%
\BDBL {}Owen, C\BPBI J.%
\end{APACrefauthors}%
\unskip\
\newblock
\APACrefYearMonthDay{2008}{jul}{}.
\newblock
{\BBOQ}\APACrefatitle {{Effects on magnetic reconnection of a density asymmetry
  across the current sheet}} {{Effects on magnetic reconnection of a density
  asymmetry across the current sheet}}.{\BBCQ}
\newblock
\APACjournalVolNumPages{Annales Geophysicae}{26}{8}{2471--2483}.
\newblock
\begin{APACrefDOI} \doi{10.5194/angeo-26-2471-2008} \end{APACrefDOI}
\PrintBackRefs{\CurrentBib}

\bibitem [\protect \citeauthoryear {%
Toledo-Redondo%
, Vaivads%
, Andr{\'{e}}%
\BCBL {}\ \BBA {} Khotyaintsev%
}{%
Toledo-Redondo%
\ \protect \BOthers {.}}{%
{\protect \APACyear {2015}}%
}]{%
Toledo-Redondo2015}
\APACinsertmetastar {%
Toledo-Redondo2015}%
\begin{APACrefauthors}%
Toledo-Redondo, S.%
, Vaivads, A.%
, Andr{\'{e}}, M.%
\BCBL {}\ \BBA {} Khotyaintsev, Y\BPBI V.%
\end{APACrefauthors}%
\unskip\
\newblock
\APACrefYearMonthDay{2015}{}{}.
\newblock
{\BBOQ}\APACrefatitle {{Modification of the Hall physics in magnetic
  reconnection due to cold ions at the Earth's magnetopause}} {{Modification of
  the Hall physics in magnetic reconnection due to cold ions at the Earth's
  magnetopause}}.{\BBCQ}
\newblock
\APACjournalVolNumPages{Geophysical Research Letters}{42}{15}{6146--6154}.
\newblock
\begin{APACrefDOI} \doi{10.1002/2015GL065129} \end{APACrefDOI}
\PrintBackRefs{\CurrentBib}

\bibitem [\protect \citeauthoryear {%
Toledo‐Redondo%
\ \protect \BOthers {.}}{%
Toledo‐Redondo%
\ \protect \BOthers {.}}{%
{\protect \APACyear {2021}}%
}]{%
ToledoRedondo2021}
\APACinsertmetastar {%
ToledoRedondo2021}%
\begin{APACrefauthors}%
Toledo‐Redondo, S.%
, Andr{\'{e}}, M.%
, Aunai, N.%
, Chappell, C\BPBI R.%
, Dargent, J.%
, Fuselier, S\BPBI A.%
\BDBL {}Vines, S\BPBI K.%
\end{APACrefauthors}%
\unskip\
\newblock
\APACrefYearMonthDay{2021}{sep}{}.
\newblock
{\BBOQ}\APACrefatitle {{Impacts of Ionospheric Ions on Magnetic Reconnection
  and Earth's Magnetosphere Dynamics}} {{Impacts of Ionospheric Ions on
  Magnetic Reconnection and Earth's Magnetosphere Dynamics}}.{\BBCQ}
\newblock
\APACjournalVolNumPages{Reviews of Geophysics}{59}{3}{}.
\newblock
\begin{APACrefDOI} \doi{10.1029/2020rg000707} \end{APACrefDOI}
\PrintBackRefs{\CurrentBib}

\bibitem [\protect \citeauthoryear {%
Walsh%
, Sibeck%
, Nishimura%
\BCBL {}\ \BBA {} Angelopoulos%
}{%
Walsh%
\ \protect \BOthers {.}}{%
{\protect \APACyear {2013}}%
}]{%
Walsh2013}
\APACinsertmetastar {%
Walsh2013}%
\begin{APACrefauthors}%
Walsh, B\BPBI M.%
, Sibeck, D\BPBI G.%
, Nishimura, Y.%
\BCBL {}\ \BBA {} Angelopoulos, V.%
\end{APACrefauthors}%
\unskip\
\newblock
\APACrefYearMonthDay{2013}{}{}.
\newblock
{\BBOQ}\APACrefatitle {{Statistical analysis of the plasmaspheric plume at the
  magnetopause}} {{Statistical analysis of the plasmaspheric plume at the
  magnetopause}}.{\BBCQ}
\newblock
\APACjournalVolNumPages{Journal of Geophysical Research: Space
  Physics}{118}{8}{4844--4851}.
\newblock
\begin{APACrefDOI} \doi{10.1002/jgra.50458} \end{APACrefDOI}
\PrintBackRefs{\CurrentBib}

\bibitem [\protect \citeauthoryear {%
R.~Wang%
\ \protect \BOthers {.}}{%
R.~Wang%
\ \protect \BOthers {.}}{%
{\protect \APACyear {2017}}%
}]{%
Wang2017}
\APACinsertmetastar {%
Wang2017}%
\begin{APACrefauthors}%
Wang, R.%
, Nakamura, R.%
, Lu, Q.%
, Baumjohann, W.%
, Ergun, R\BPBI E.%
, Burch, J\BPBI L.%
\BDBL {}Wang, S.%
\end{APACrefauthors}%
\unskip\
\newblock
\APACrefYearMonthDay{2017}{}{}.
\newblock
{\BBOQ}\APACrefatitle {{Electron-Scale Quadrants of the Hall Magnetic Field
  Observed by the Magnetospheric Multiscale spacecraft during Asymmetric
  Reconnection}} {{Electron-Scale Quadrants of the Hall Magnetic Field Observed
  by the Magnetospheric Multiscale spacecraft during Asymmetric
  Reconnection}}.{\BBCQ}
\newblock
\APACjournalVolNumPages{Physical Review Letters}{118}{17}{1--5}.
\newblock
\begin{APACrefDOI} \doi{10.1103/PhysRevLett.118.175101} \end{APACrefDOI}
\PrintBackRefs{\CurrentBib}

\bibitem [\protect \citeauthoryear {%
S.~Wang%
, Kistler%
, Mouikis%
, Liu%
\BCBL {}\ \BBA {} Genestreti%
}{%
S.~Wang%
\ \protect \BOthers {.}}{%
{\protect \APACyear {2014}}%
}]{%
Wang2014}
\APACinsertmetastar {%
Wang2014}%
\begin{APACrefauthors}%
Wang, S.%
, Kistler, L\BPBI M.%
, Mouikis, C\BPBI G.%
, Liu, Y.%
\BCBL {}\ \BBA {} Genestreti, K\BPBI J.%
\end{APACrefauthors}%
\unskip\
\newblock
\APACrefYearMonthDay{2014}{}{}.
\newblock
{\BBOQ}\APACrefatitle {{Hot magnetospheric O+ and cold ion behavior in
  magnetopause reconnection: Cluster observations}} {{Hot magnetospheric O+ and
  cold ion behavior in magnetopause reconnection: Cluster
  observations}}.{\BBCQ}
\newblock
\APACjournalVolNumPages{Journal of Geophysical Research: Space
  Physics}{119}{12}{9601--9623}.
\newblock
\begin{APACrefDOI} \doi{10.1002/2014JA020402} \end{APACrefDOI}
\PrintBackRefs{\CurrentBib}

\bibitem [\protect \citeauthoryear {%
S.~Wang%
, Kistler%
, Mouikis%
\BCBL {}\ \BBA {} Petrinec%
}{%
S.~Wang%
\ \protect \BOthers {.}}{%
{\protect \APACyear {2015}}%
}]{%
Wang2015}
\APACinsertmetastar {%
Wang2015}%
\begin{APACrefauthors}%
Wang, S.%
, Kistler, L\BPBI M.%
, Mouikis, C\BPBI G.%
\BCBL {}\ \BBA {} Petrinec, S\BPBI M.%
\end{APACrefauthors}%
\unskip\
\newblock
\APACrefYearMonthDay{2015}{aug}{}.
\newblock
{\BBOQ}\APACrefatitle {{Dependence of the dayside magnetopause reconnection
  rate on local conditions}} {{Dependence of the dayside magnetopause
  reconnection rate on local conditions}}.{\BBCQ}
\newblock
\APACjournalVolNumPages{Journal of Geophysical Research: Space
  Physics}{120}{8}{6386--6408}.
\newblock
\begin{APACrefURL}
  \url{https://onlinelibrary.wiley.com/doi/10.1002/2015JA021524}
  \end{APACrefURL}
\newblock
\begin{APACrefDOI} \doi{10.1002/2015JA021524} \end{APACrefDOI}
\PrintBackRefs{\CurrentBib}

\bibitem [\protect \citeauthoryear {%
Yamada%
, Kulsrud%
\BCBL {}\ \BBA {} Ji%
}{%
Yamada%
\ \protect \BOthers {.}}{%
{\protect \APACyear {2010}}%
}]{%
Yamada2010}
\APACinsertmetastar {%
Yamada2010}%
\begin{APACrefauthors}%
Yamada, M.%
, Kulsrud, R.%
\BCBL {}\ \BBA {} Ji, H.%
\end{APACrefauthors}%
\unskip\
\newblock
\APACrefYearMonthDay{2010}{}{}.
\newblock
{\BBOQ}\APACrefatitle {{Magnetic reconnection}} {{Magnetic
  reconnection}}.{\BBCQ}
\newblock
\APACjournalVolNumPages{Reviews of Modern Physics}{82}{1}{603--664}.
\newblock
\begin{APACrefDOI} \doi{10.1103/RevModPhys.82.603} \end{APACrefDOI}
\PrintBackRefs{\CurrentBib}

\bibitem [\protect \citeauthoryear {%
Young%
\ \protect \BOthers {.}}{%
Young%
\ \protect \BOthers {.}}{%
{\protect \APACyear {2016}}%
}]{%
Young2016}
\APACinsertmetastar {%
Young2016}%
\begin{APACrefauthors}%
Young, D\BPBI T.%
, Burch, J\BPBI L.%
, Gomez, R\BPBI G.%
, {De Los Santos}, A.%
, Miller, G\BPBI P.%
, Wilson, P.%
\BDBL {}Webster, J\BPBI M.%
\end{APACrefauthors}%
\unskip\
\newblock
\APACrefYearMonthDay{2016}{}{}.
\newblock
{\BBOQ}\APACrefatitle {{Hot Plasma Composition Analyzer for the Magnetospheric
  Multiscale Mission}} {{Hot Plasma Composition Analyzer for the Magnetospheric
  Multiscale Mission}}.{\BBCQ}
\newblock
\APACjournalVolNumPages{Space Science Reviews}{199}{1-4}{407--470}.
\newblock
\begin{APACrefURL} \url{http://dx.doi.org/10.1007/s11214-014-0119-6}
  \end{APACrefURL}
\newblock
\begin{APACrefDOI} \doi{10.1007/s11214-014-0119-6} \end{APACrefDOI}
\PrintBackRefs{\CurrentBib}

\bibitem [\protect \citeauthoryear {%
Zhang%
\ \protect \BOthers {.}}{%
Zhang%
\ \protect \BOthers {.}}{%
{\protect \APACyear {2018}}%
}]{%
Zhang2018}
\APACinsertmetastar {%
Zhang2018}%
\begin{APACrefauthors}%
Zhang, Q.%
, Lockwood, M.%
, Foster, J\BPBI C.%
, Zong, Q.%
, Dunlop, M\BPBI W.%
, Zhang, S.%
\BDBL {}Zhang, B.%
\end{APACrefauthors}%
\unskip\
\newblock
\APACrefYearMonthDay{2018}{}{}.
\newblock
{\BBOQ}\APACrefatitle {{Observations of the step-like accelerating processes of
  cold ions in the reconnection layer at the dayside magnetopause}}
  {{Observations of the step-like accelerating processes of cold ions in the
  reconnection layer at the dayside magnetopause}}.{\BBCQ}
\newblock
\APACjournalVolNumPages{Science Bulletin}{63}{1}{31--37}.
\newblock
\begin{APACrefURL} \url{https://doi.org/10.1016/j.scib.2018.01.003}
  \end{APACrefURL}
\newblock
\begin{APACrefDOI} \doi{10.1016/j.scib.2018.01.003} \end{APACrefDOI}
\PrintBackRefs{\CurrentBib}

\end{thebibliography}

%
%
%
%
\begin{figure}
 \includegraphics[scale=0.45]{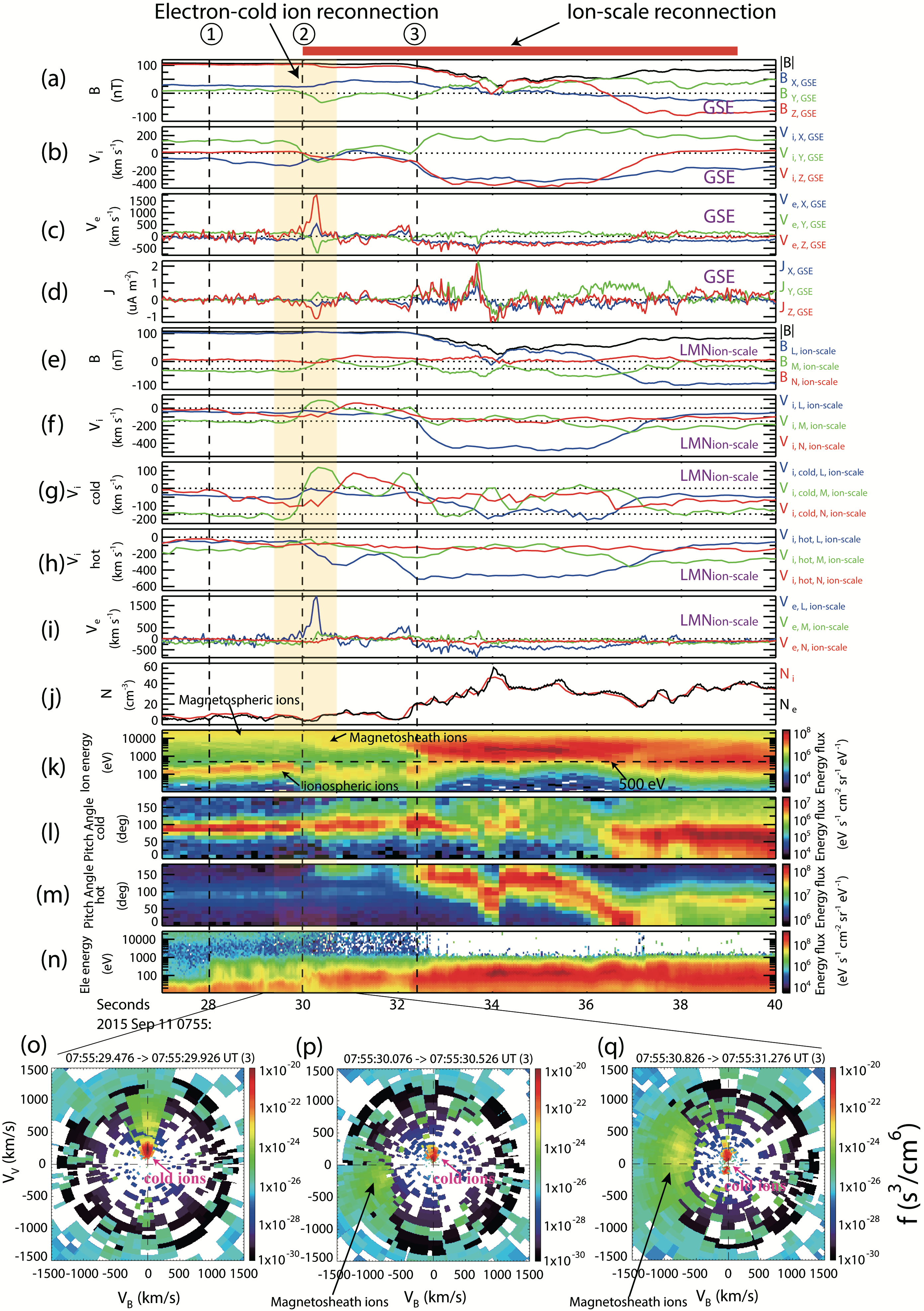}
 \caption{MMS1 spacecraft-frame observations of a multiscale, multi-type reconnection at the dayside magnetopause on 11 September 2015. (a) magnetic field in GSE; (b) ion and (c) electron bulk velocity in GSE; (d) plasma current density in GSE; (e) magnetic field in $[LMN]_{ion-scale}$; (f) ion and (i) electron bulk velocity in $[LMN]_{ion-scale}$; (g) cold and (h) hot ion bulk velocity in $[LMN]_{ion-scale}$; (j) plasma density; (k) ion and (n) electron spectrograms. The horizontal dashed line marks 500 eV in Figure 1k. (l) cold and (m) hot ion pitch angle distributions; (o-q) $V_B-V_V$ cuts of FPI ion distributions at 0.45 s interval during 07:55:29.476-07:55:31.276 UT, where $V_B$ is along the magnetic field, and $V_V$ is along $(\textbf{v}\times \textbf{b})\times \textbf{b}$ ($\textbf{b}$ and $\textbf{v}$ are unit vectors of the magnetic field and ion bulk velocity). The red bar at the top of Figure 1a marks the ion-scale reconnection. The yellow-shaded region marks the electron-cold ion current sheet. Vertical dashed lines denote flow boundaries: \circled{1} Electron boundary relates to the inner edge of magnetosheath electrons; \circled{2} Hot ion boundary relates to the inner edge of magnetosheath ions; \circled{3} Cold ion boundary relates to the outer edge of cold ions of ionospheric origin. MMS = Magnetospheric Multiscale mission. GSE = Geocentric Solar Ecliptic coordinates. $[LMN]_{ion-scale}$ = Local Magnetic Normal coordinates of the ion-scale current sheet. Cold-ion energy range: 10-500 eV; hot-ion energy range: 500-30000 eV.}
 \end{figure}

 \begin{figure}
 \includegraphics[scale=0.3]{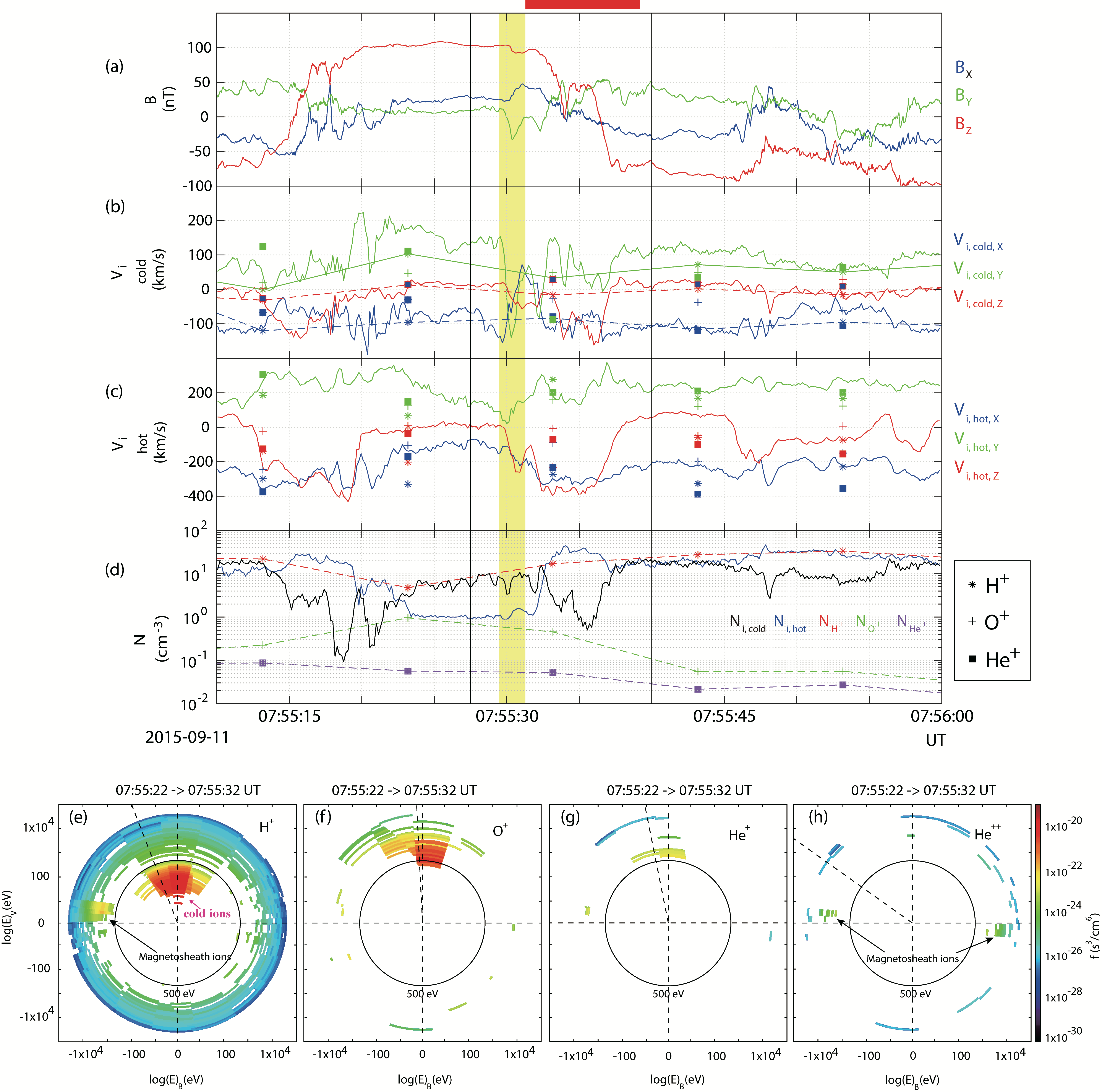}
 \caption{The comparison between FPI and HPCA measurements from MMS1 in GSE. (a) magnetic field; (b) cold and (c) hot ion bulk velocity; (d) ion density. The stars, pluses, and squares denote $H^+$, $O^+$, and $He^+$, respectively. (e-h) $V_B-V_V$ cuts of HPCA distributions for $H^+$,  $O^+$, $He^+$, and $He^{++}$ in the spacecraft frame during 07:55:22-07:55:32 UT. The red bar at the top of panel (a) and the yellow-shaded region mark the ion-scale reconnection and electron-cold ion current sheet, respectively (the same as Figure 1).}
 \end{figure}

 \begin{figure}
 \includegraphics[scale=0.15]{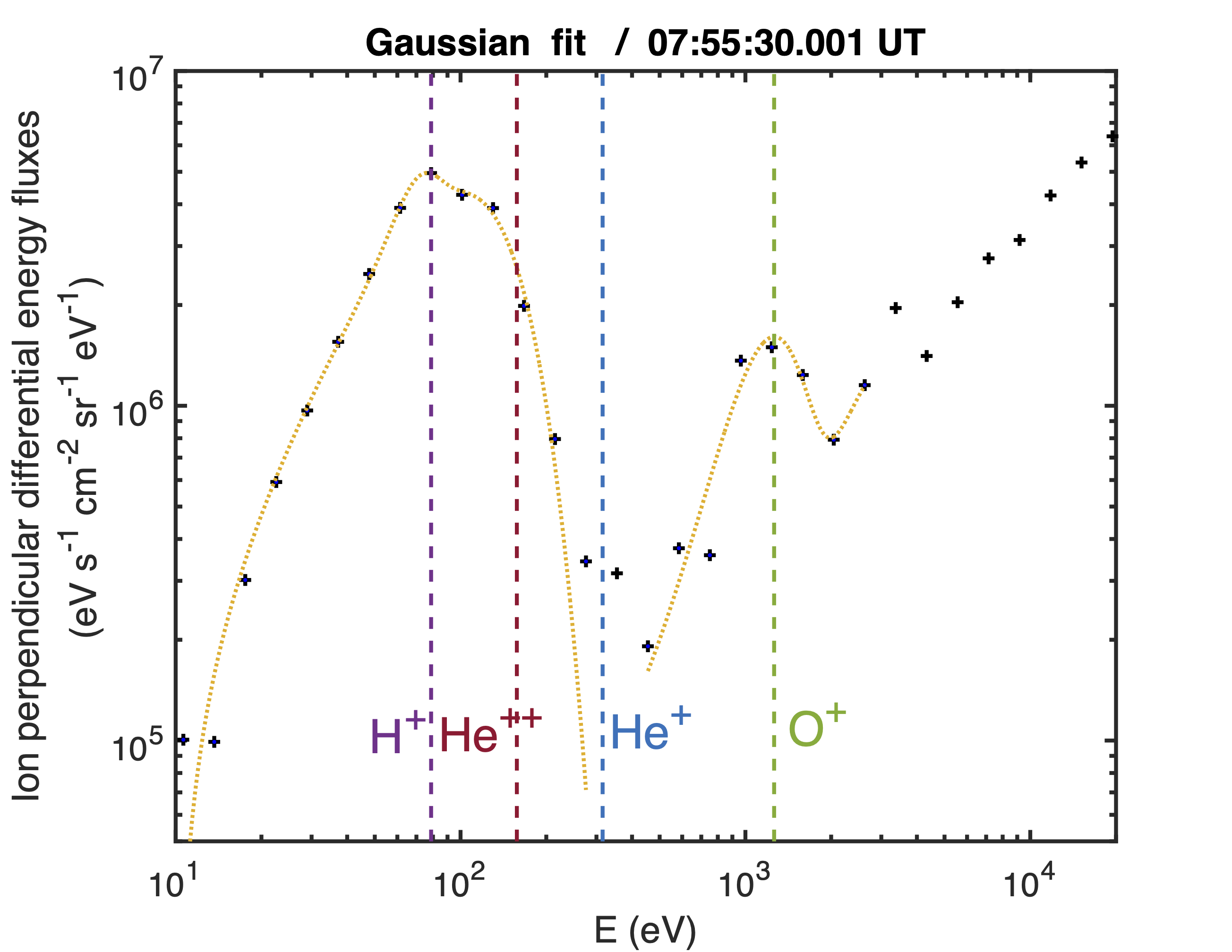}
 \caption{Ion species separation analysis based on the measurements of FPI onboard MMS1 at 07:55:30.001 UT. Two Gaussian curves (orange dotted curve) are used to fit the one-dimensional cut of the ion perpendicular energy fluxes (pluses) in the spacecraft frame. The purple, red, blue, and green vertical lines denote the predicted energies of $H^+$, $He^{++}$, $He^+$, and $O^+$ at energy flux peaks, respectively.}
 \end{figure}

 \begin{figure}
 \includegraphics[scale=0.8]{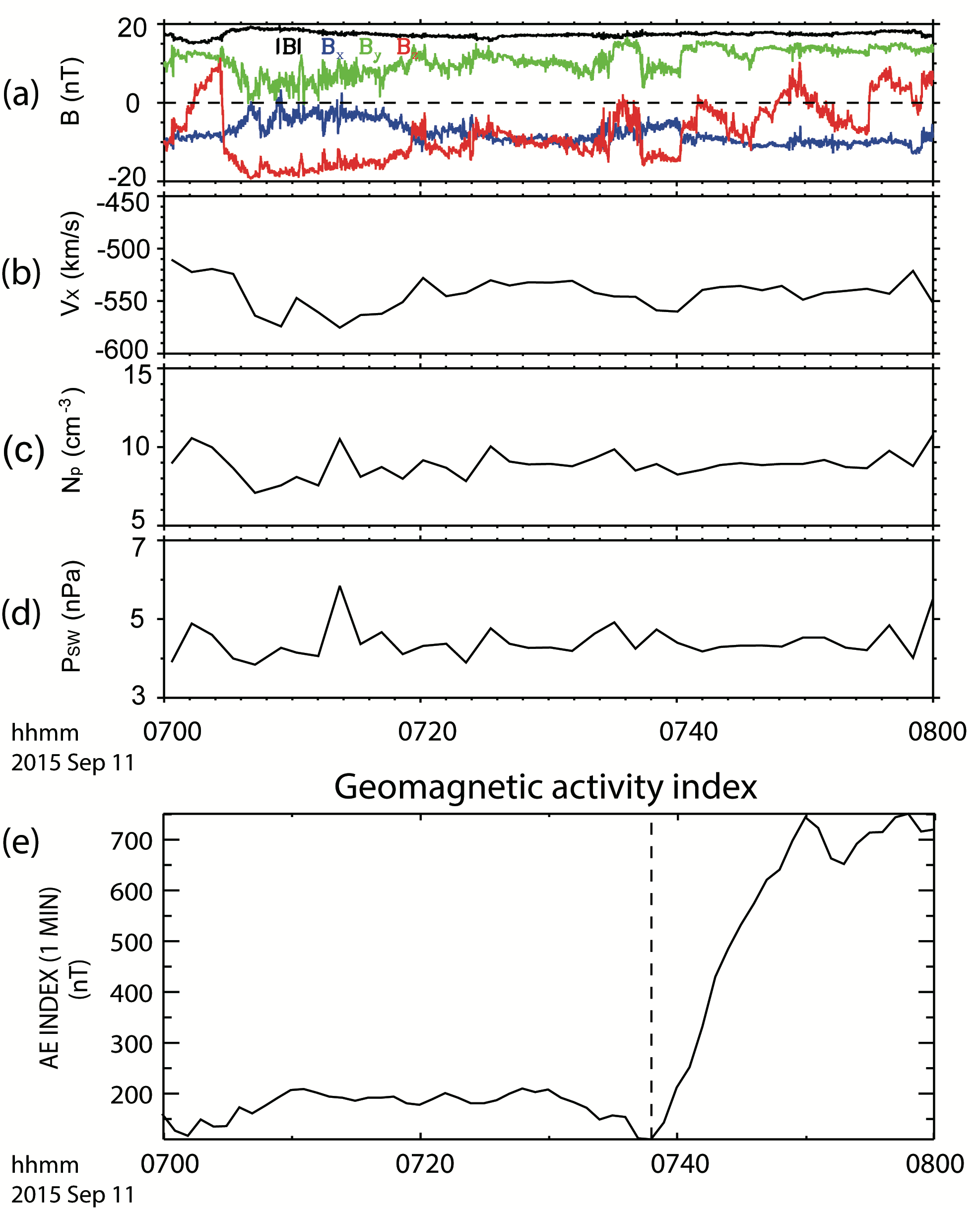}
 \caption{Data from 07:00:00 UT to 08:00:00 UT on 11 September 2015. (a-d) Solar wind conditions from the Wind spacecraft in GSE. (a) interplanetary magnetic field; (b) plasma velocity $V_{X,GSE}$ component; (c) proton density; (d) solar wind dynamic pressure $P_{SW}$; (e) AE index from the NASA OMNIWeb. The vertical dashed line is at 07:38:00 UT.}
 \end{figure}
 
 \begin{figure}
 \includegraphics[scale=0.25]{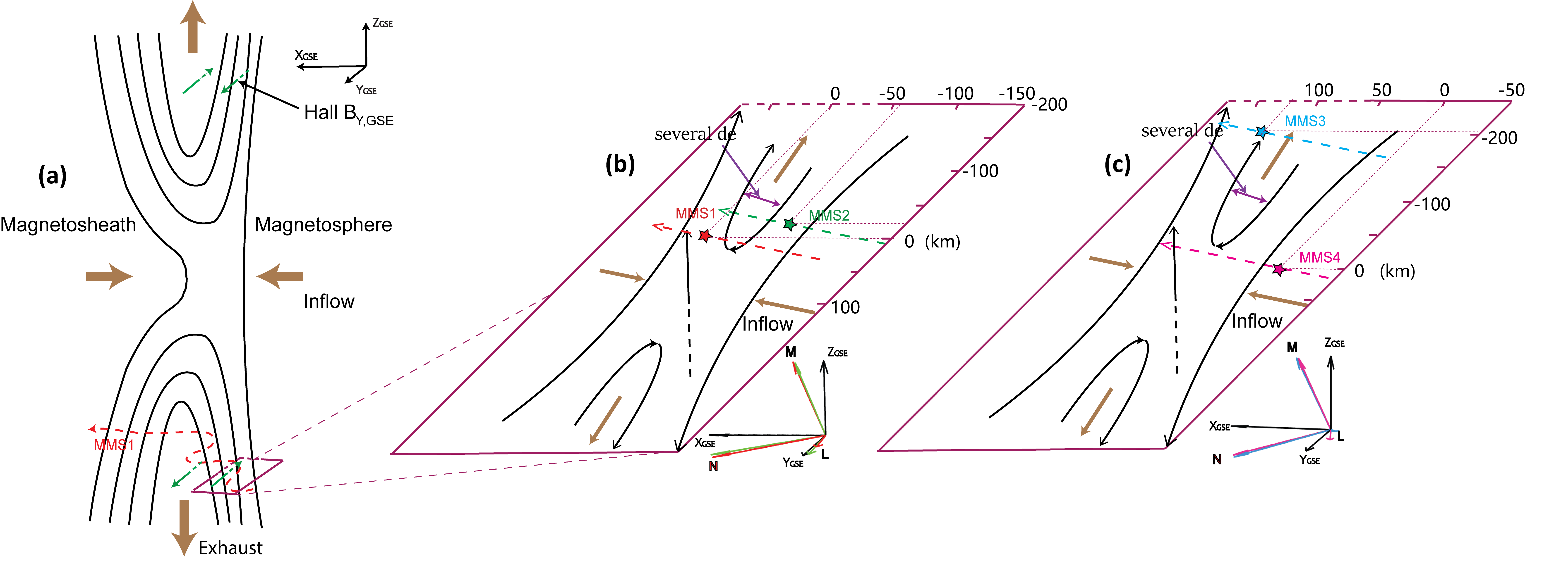}
 \caption{A schematic illustration for the multiscale, multi-type reconnection. (a) An ion-scale reconnection primarily in $XZ_{GSE}$ plane. The purple box in $XY_{GSE}$ plane marks the electron-cold ion current sheet located at the edge of the ion-scale reconnection. The green arrows mark the out-of-plane Hall magnetic field $B_{Y,GSE}$. The red dashed curve marks the MMS1 trajectory. (b) zoom-in electron-cold ion reconnection at $\sim$07:55:30.4 UT, where MMS1 is located at the origin of the coordinate (MMS1 went across the electron jet, and MMS2 begins traversing). (c) the same electron-cold ion current sheet at $\sim$07:55:33.4 UT, where MMS4 is located at the origin of the coordinate (MMS3 observes the electron jet and MMS4 has not started yet). The coordinate scale is in units of km. The red, green, blue, and pink dashed lines mark MMS1, 2, 3, and 4 trajectories.}
 \end{figure}

 \begin{figure}
 \includegraphics[scale=0.35]{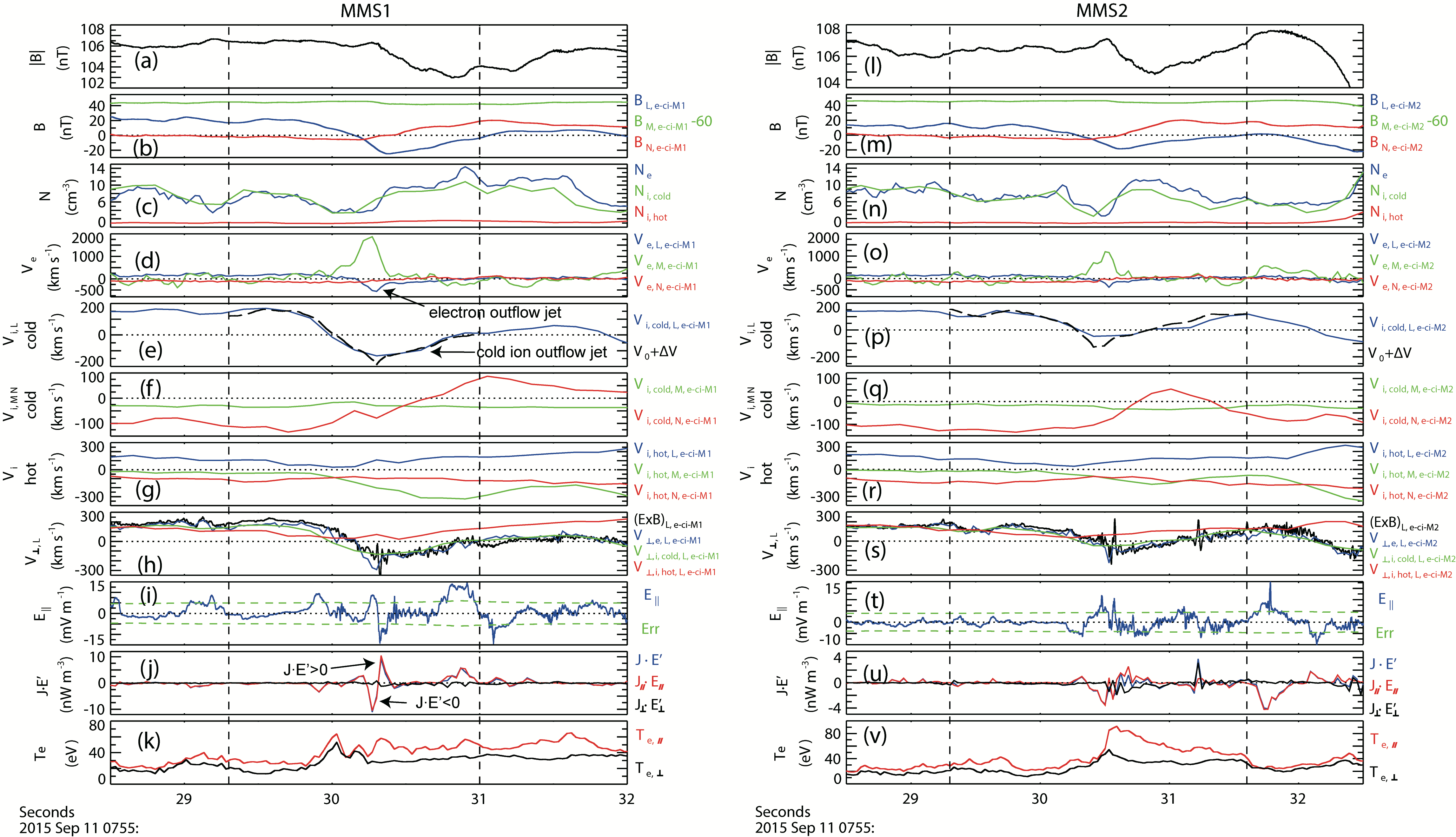}
 \caption{MMS1 and MMS2 observations of the electron-cold ion current sheet. Data are displayed in $[LMN]_{e-ci-M1}$ and $[LMN]_{e-ci-M2}$ coordinates, respectively. (a, l) magnetic field magnitude; (b, m) magnetic field components; (c, n) plasma density; (d, o) electron bulk velocity; (e-f, p-q) cold ion bulk velocity; (g, r) hot ion bulk velocity; (h, s) comparison of ion and electron perpendicular velocities and $\frac{\textbf{E}\times \textbf{B}}{B^2}$ in $L_{e-ci-*}$ direction; (i, t) parallel electric field smoothed with a time window of 0.01 s. The green dashed lines represent the electric field errors. (j, u) energy conversion rate: $\textbf{J}\cdot \textbf{E}^{'} = \textbf{J}\cdot (\textbf{E}+ \textbf{v}_e\times \textbf{B})$,  $\textbf{J}_\parallel \cdot \textbf{E}_\parallel$, and $\textbf{J}_{\perp} \cdot \textbf{E}_{\perp}^{'}$. (k, v) electron temperature.$[LMN]_{e-ci-M1}$ eigenvectors: $\textbf{L}_{e-ci-M1}$ = $[-0.12, 0.99, 0.12]_{GSE}$, $\textbf{M}_{e-ci-M1} = [0.28, -0.08, 0.96]_{GSE}$, and $\textbf{N}_{e-ci-M1} = [0.95, 0.14, -0.26]_{GSE}$. $[LMN]_{e-ci-M2}$ eigenvectors: $\textbf{L}_{e-ci-M2} = [-0.05, 0.996, 0.02]_{GSE}$, $\textbf{M}_{e-ci-M2} = [0.25, -0.01, 0.97]_{GSE}$, and $\textbf{N}_{e-ci-M2} = [0.98, 0.05, -0.25]_{GSE}$.}
 \end{figure}

 \begin{figure}
 \includegraphics[scale=0.35]{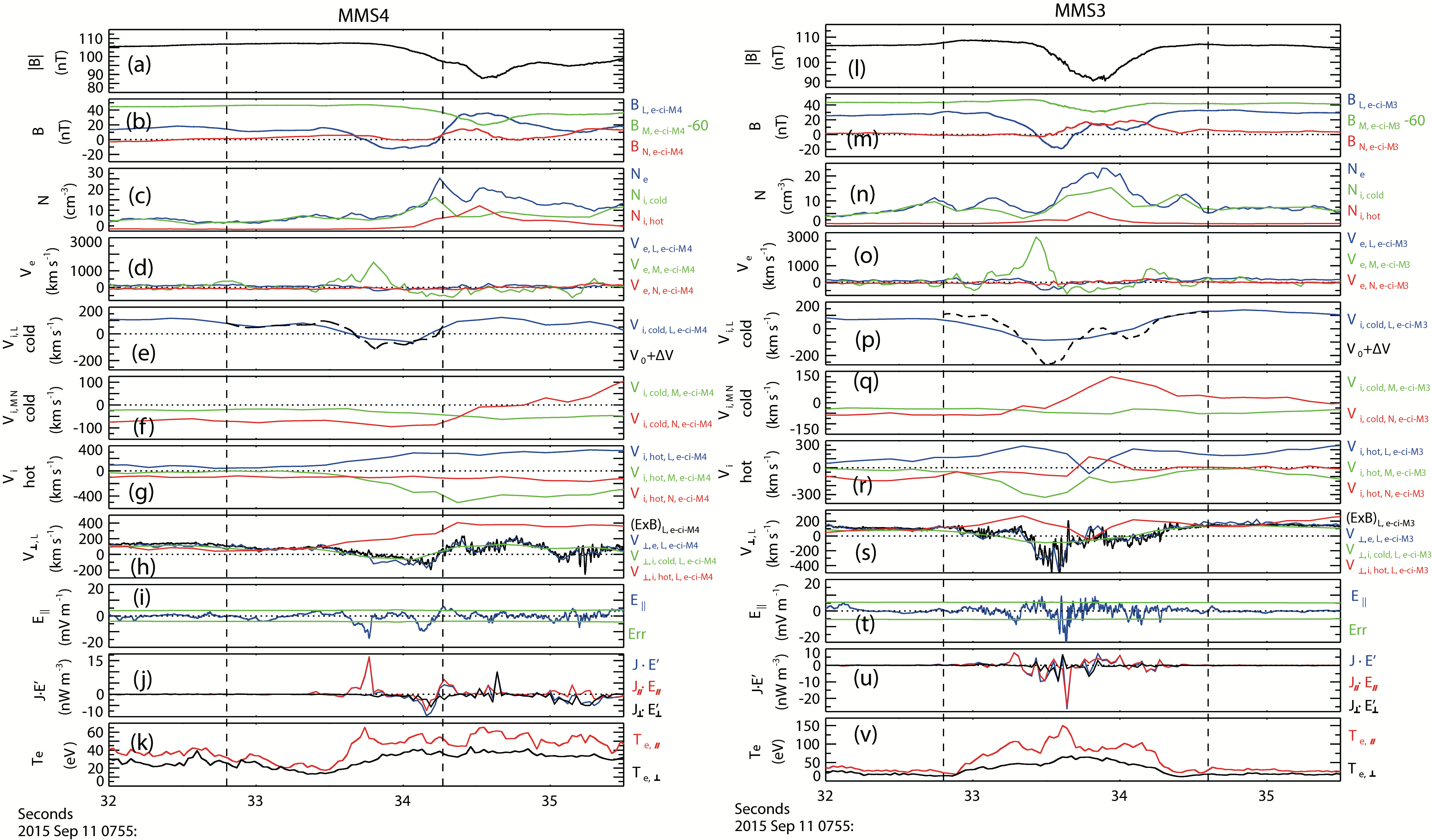}
 \caption{MMS4 and MMS3 observations of the electron-cold ion current sheet. Same format as Figure 6. $[LMN]_{e-ci-M4}$ eigenvectors: $\textbf{L}_{e-ci-M4} = [-0.23, 0.95, 0.15]_{GSE}$, $\textbf{M}_{e-ci-M4} = [0.34, -0.07, 0.93]_{GSE}$, and $\textbf{N}_{e-ci-M4} = [0.91, 0.27, -0.31]_{GSE}$. $[LMN]_{e-ci-M3}$ eigenvectors: $\textbf{L}_{e-ci-M3} = [-0.28, 0.92, 0.26]_{GSE}$, $\textbf{M}_{e-ci-M3} = [0.38, -0.15, 0.91]_{GSE}$, and $\textbf{N}_{e-ci-M3} = [0.88, 0.36, -0.31]_{GSE}$.}
 \end{figure}
 
  \begin{figure}
 \includegraphics[scale=0.32]{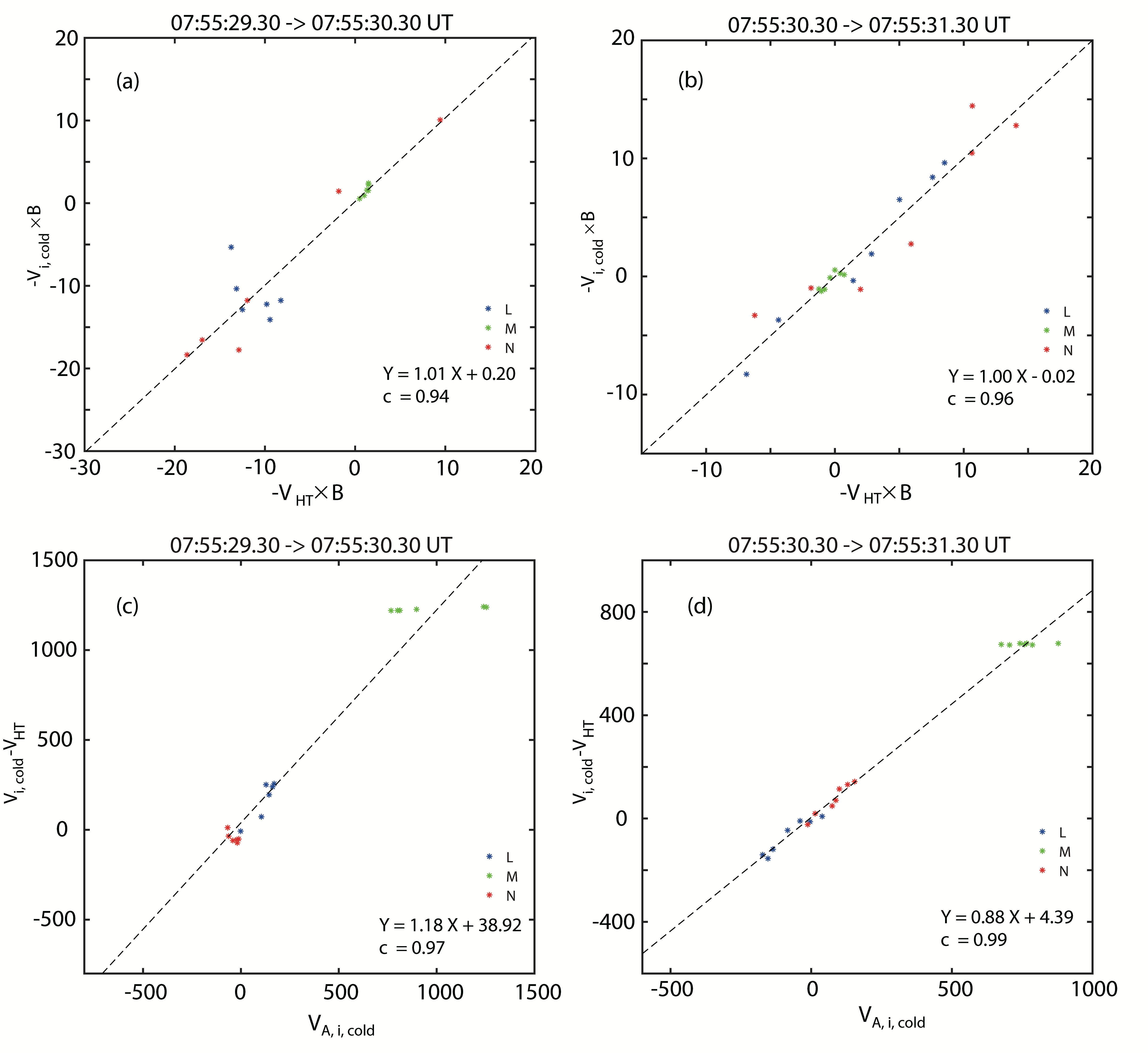}
 \caption{Walen relations on both sides of the electron-cold ion current sheet observed by MMS1. The blue, green, and red dots denote the L-, M-, and N- components in $[LMN]_{e-ci-M1}$, respectively. (a-b) the correlations on a point-by-point basis between cold-ion convection electric fields and $-\textbf{V}_{HT}\times \textbf{B}$; (c-d) the correlations between the local cold-ion Alfvén velocity ($V_{A,i,cold}=B/\sqrt{4\pi m_{i}n_{i,cold}}$) and cold-ion bulk velocity in the DeHoffmann-Teller frame. The parameter c is the linear correlation coefficient.}
 \end{figure}

\end{document}